\documentclass[%
reprint,
nofootinbib,
amsmath,amssymb,
aps,
]{revtex4-1}

\usepackage{color}
\usepackage{textcomp}
\usepackage{hyperref}
\usepackage{graphicx}
\usepackage{dcolumn}
\usepackage{bm}

\newcommand{\beginsupplement}{%
	\setcounter{table}{0}
	\renewcommand{\thetable}{S\arabic{table}}%
	\setcounter{figure}{0}
	\renewcommand{\thefigure}{S\arabic{figure}}%
}

\usepackage{tikz}
\usepackage{hyperref}

\begin{document}
	
	\preprint{APS/123-QED}
	


     \title{Modular boundaries in recurrent neural networks}


	\author{Jacob C. Tanner$^{1,2}$}
        \author{Sina Mansour L.$^{3}$}
         \author{Ludovico Coletta$^4$}
        \author{Alessandro Gozzi$^5$}
        \author{Richard F. Betzel$^{1,2,6}$}
        \email{rbetzel @ indiana.edu}

    \affiliation{
    $^1$Cognitive Science Program, $^2$ School of Informatics, Computing, and Engineering, Indiana University, Bloomington, IN 47405 \\ 
    $^3$ National University of Singapore\\
    $^4$Fondazione Bruno Kessler, Trento, Trento, \\ 
    $^5$Functional Neuroimaging Lab, Istituto Italiano di Tecnologia, Center for Neuroscience and Cognitive Systems, Rovereto, Italy,\\
    $^6$Department of Psychological and Brain Sciences, Indiana University, Bloomington, IN 47405}
	
	\date{\today}
	\begin{abstract}

Recent theoretical and experimental work in neuroscience has focused on the representational and dynamical character of neural manifolds --subspaces in neural activity space wherein many neurons coactivate. Importantly, neural populations studied under this ``neural manifold hypothesis'' are continuous and not cleanly divided into separate neural populations. This perspective clashes with the ``modular hypothesis'' of brain organization, wherein neural elements maintain an ``all-or-nothing'' affiliation with modules. In line with this modular hypothesis, recent research on recurrent neural networks suggests that multi-task networks become modular across training, such that different modules specialize for task-general dynamical motifs \cite{yang2019task,driscoll2024flexible}. If the modular hypothesis is true, then it would be important to use a dimensionality reduction technique that captures modular structure. Here, we investigate the features of such a method. We leverage RNNs as a model system to study the character of \emph{modular} neural populations, using a community detection method from network science known as \emph{modularity maximization} to partition neurons into distinct modules. These partitions allow us to ask the following question: \emph{do these modular boundaries matter to the system?} We find evidence that they do. First, we find that these boundaries neatly divide the representational content and role of neurons. Next, we find that these boundaries can be directly inferred from features of the weight matrix in feed-forward neural networks, and are related to clustering of the Jacobian matrix in recurrent neural networks. We also find that the weights of input neurons to recurrent neurons partially inform their modular structure, an observation that we corroborated using structural and functional imaging data from mice and humans. Finally, we find that the dynamics of these RNNs reflected the boundaries between modules. Collectively, our results suggest that neural populations in RNNs sometimes form modules, and that the boundaries between such modules -- as defined by a network science based dimensionality reduction technique -- are informative about the systems behavior. Our work invites new avenues for the analysis of low-dimensional dynamics from a modular perspective.

	\end{abstract}
	
	\maketitle
	\section*{Introduction}

In recent decades, technological advances in multi-unit recording have revealed that neural activity is often low-dimensional, reflected by the tendency for neurons to coactivate \cite{cunningham2014dimensionality,kipke2008advanced,kerr2008imaging,ahrens2013whole}. This has led to a proliferation of research on the functional significance of low-dimensional neural manifolds, often emphasizing that these manifolds carry task-relevant information like choice information, or other task-related variables, and that neural dynamics are responsible for transforming input into this task-relevant form \cite{ebitz2021population,saxena2019towards,mante2013context,cunningham2014dimensionality}. 

While there are many methods for studying low-dimensional neural manifolds, by far the most commonly used method is principal component analysis (PCA). PCA finds directions in state space along which many neurons tend to coactivate and treats those directions as new basis vectors in a lower-dimensional state space. While this method has its advantages for the analysis of neural populations, the use of different tools often affords novel questions. Implicit in the use of PCA as a tool is the idea that neurons do not discretely separate into sub-populations. Instead, neurons jointly participate in lower-dimensional manifolds. For this reason, we argue that neural manifolds are causally opaque in the following sense: it is unclear what manipulations we can perform to learn more about the systems causal relationships. 

In parallel, many researchers have advocated for concepts like cellular assemblies and neural ensembles, concepts that implicitly or explicitly align with the ``modular hypothesis'' wherein neurons maintain an ``all-or-nothing'' affiliation with modules (e.g. \cite{buzsaki2010neural,carrillo2016imprinting,carrillo2019controlling,yuste2024neuronal}). Whereas PCA finds manifolds, cellular assemblies and neuronal ensembles focus on \emph{distinct} and \emph{modular} populations of neurons. In other words, neurons belong to one and only one population/module. Consequently, we are led to inquire: Are these modular boundaries between populations meaningful to the system itself \cite{gyorgy2019brain}? This is a natural question to ask of modular systems, and as such modular systems constitutionally afford a number of possible causal manipulation strategies. For example, how do the effects of lesions to one module differ from the effect of lesions to another module?

One potentially fruitful strategy for investigating the importance of modular boundaries between neuronal groups is using \emph{in silico} simulations of neurons that have been trained to accomplish a task or a goal. Recurrent neural networks offer trainable systems that operate based on a network of dynamically interacting parts and as such are ideal models for investigating such questions \cite{yang2020artificial,lindsay2022testing,perich2020rethinking,beer1995dynamics,beer2003dynamics,beer1996toward,finkelstein2021attractor,sussillo2014neural,mante2013context}. Not only do these artificial systems provide us with complete access to all the information that is important for their function, but they also offer a safe and ethically neutral platform for causal manipulation. 

Indeed, recent research used causal manipulation (e.g. artificial lesions to neurons within a module) to study RNN models trained to perform multiple tasks and found that these networks organize themselves into distinct modules specialized for task-general dynamical motifs \cite{yang2019task,driscoll2024flexible}, lesions to which had isolated effects on each related task set, or dynamical motif. In fact, there is a rich history of literature investigating the modular hypothesis in similar models. For example, modularity has been shown to emerge in the present of multiple/hierarchical goals \cite{kashtan2005spontaneous,espinosa2010specialization}. Other studies have found that modularity can emerge for ``free'' from connection costs \cite{clune2013evolutionary}, and distance constraints \cite{achterberg2022spatially}. Still other studies have shown that the modularity that emerges from simple cost and distance constraints, nonetheless yields distinctive functions/advantages, such as assisting the network in remembering old skills while acquiring new ones \cite{ellefsen2015neural}, as well as specializing locally clustered nodes for a similar purpose \cite{achterberg2022spatially}. Other studies have forced modularity in the physical connections between units and found that modules can assist with memory tasks \cite{rodriguez2019optimal}, information filtration \cite{kleinman2021mechanistic}, and enable the separation of dynamic time-scales \cite{pan2009modularity}.

However the methods used to identify modules in these papers, while important for investigating the modular hypothesis generally, do not qualify as dimensionality reduction techniques given that they do not reduce the number of dimensions for the system while preserving as much of the original information as possible. The network science tool that we explore here, when applied to a correlation matrix (also known in some research programs as a ``functional network'') has two important functions: (1) it produces a partition of neural elements into distinct modules separated by a clear boundaries between neural sub-populations and (2) it can be used to reduce the dimensionality of the system (by taking the mean activity of neural units in each module). These two functions map onto the two complimentary purposes of this paper, which are: (1) to investigate a dimensionality reduction technique that yields distinct modules, and (2) to explore the modular hypothesis by investigating the nature of the modular boundaries offered by this method.

In this work, we use RNNs trained on systems neuroscience tasks \cite{molano2022neurogym} alongside a network science tool called ``modularity maximization'' \cite{esfahlani2021modularity} to demonstrate that modular boundaries between neural populations are meaningful to the system in several important ways. First, these boundaries are relevant to the representations of the sub-populations. That is, modules can be found that divide the representations of different task-relevant variables. In the case of multi-task networks, we found that we could identify the task by the modular boundaries of time-varying networks. Second, we provide preliminary evidence that these boundaries have specific structural/connectivity-based origins. For example, we show that -- in non-recurrent feed-forward neurons -- the weights converging on a post-synaptic neuron determine the module that it belongs to. We also demonstrate a relationship between the modules in RNNs and the partial derivatives of recurrent neurons.

Finally, and perhaps most importantly, we demonstrate that these modular boundaries are important to the systems dynamics. We show that task-relevant dynamics for transforming inputs into latent variables reflect the boundaries projected onto the recurrent layer by the feed-forward input neurons. Further, we use a lesioning analysis to show that neurons in different modules contribute to system dynamics in unique ways. Finally, we develop a nullcline approximation technique to further demonstrate that system dynamics reflect modular boundaries.

Ultimately, the question of whether low-dimensional neural dynamics are modular or not modular might not be well-posed (as they might behave in a modular fashion in some cases, and not others), but by using alternative methods for studying low-dimensional dynamics different features stand out as worthy of study. PCA, as we outline further in the next section, does not create boundaries between distinct sets of neurons. By using a method that does offer modular boundaries, we can ask different questions about the representations, origins and dynamics of low-dimensional neural trajectories.

Finally, while recent theoretical and experimental work in neuroscience has focused on the neural manifold as the primary explanation for cognition and behavior, the use of causal and circuit-based descriptions of behavior remain an important venue for understanding neural systems. Our work suggests that by pairing the intuitive power of neural manifolds with dimensionality reduction methods that define modular boundaries between neural populations (e.g. modularity maximization), we might be able to more clearly uncover the neuronal circuit-level purpose, origin, and dynamics of low-dimensional population activity.

\begin{figure*}[!t]
    \centering
    \includegraphics[width=1\textwidth]{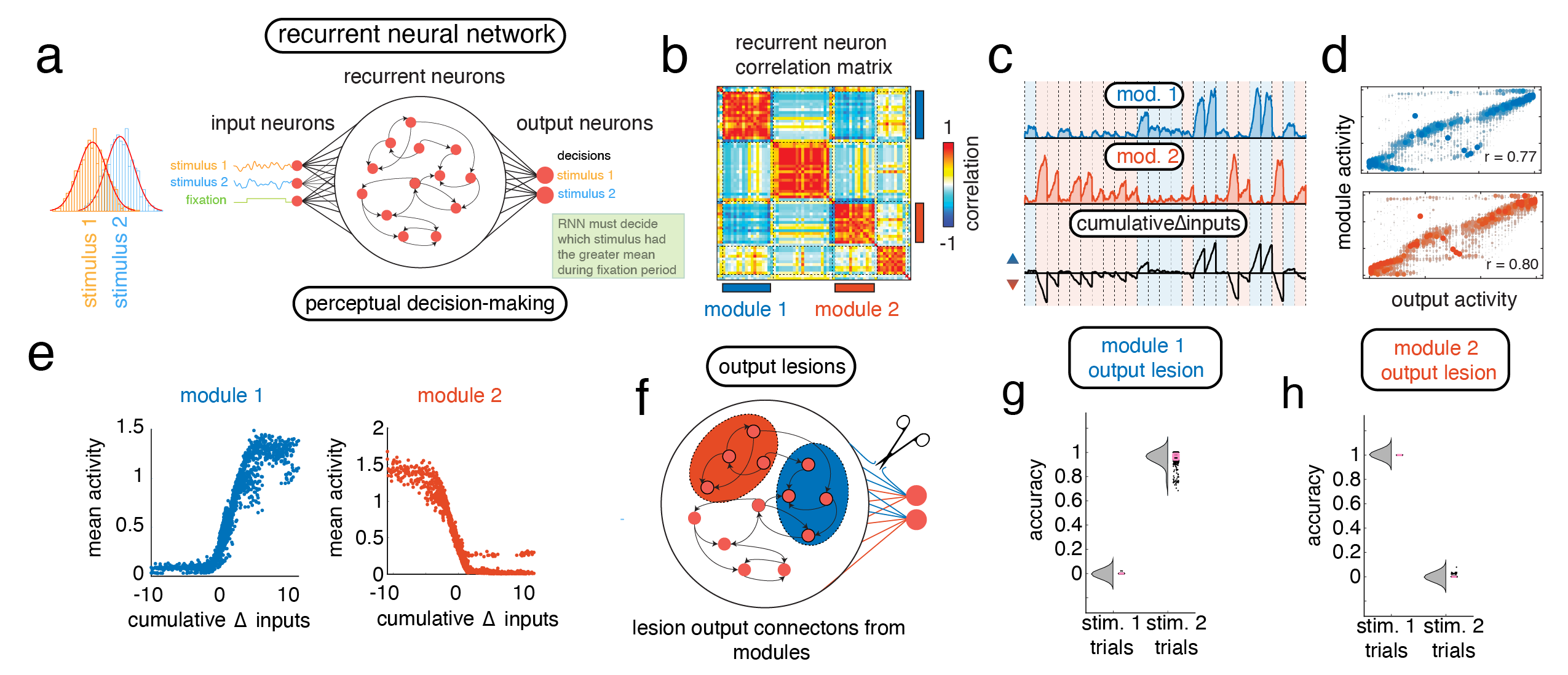}
	\caption{\textbf{Representations/selectivity profiles reflect module boundaries}, (\emph{a}) Schematic describing the perceptual decision-making task and the architecture of the RNN trained to perform it. The input neurons are given stimulus information and a fixation input. The stimuli come from two distributions with different means. The RNN must determine which of the two stimuli come from the distribution with the greater mean while the fixation input has a value of 1. When the fixation input is zero, the decision is made based on which of the output neurons have the greater activity. (\emph{b}) Correlaton matrix of recurrent neuronal activity during task trials reorganized according to modules. Modularity maximization found four modules, two of which are labeled 'mod. 1' and 'mod. 2'. The other two modules are associated with the fixation input (one activates at the beginning of the fixation period, and the other activates at the end of the fixation period). (\emph{c}) Mean activity of each population for different fixation periods as well as cumulative $\Delta$ inputs between the stimuli. Notice how the mean activity in each module tracks with this value. (\emph{d}) Two plots showing the correlation between activity of the modules and the activity of output neurons.(\emph{e}) Two plots showing that modules 1 and 2 track cases where cumulative $\Delta$ inputs is greater than 0 and less than zero respectively. (\emph{f}) Schematic showing the process of lesioning the outgoing connections from modules. (\emph{g/h}) Two sets of boxplots showing the accuracy on different trials following an lesions to module outputs. Accuracy is considered separately for trials where stimulus 1 had the higher mean than stimulus 2, and vice versa.} 
	\label{mod_func}
\end{figure*}

\section*{Principal component analysis \emph{versus} modularity maximization}

Principal Component Analysis (PCA) is a widely used dimensionality reduction technique in data analysis and machine learning and has been particularly useful in exploring lower-dimensional dynamics in neural population data. This is because PCA can transform neural activity data into a new coordinate system where the greatest variance comes to lie on the first coordinate (called the first principal component), the second greatest variance on the second coordinate, and so on. PCA starts by constructing the covariance matrix of the data, which will capture the pairwise covariance between the time courses of different neurons. The eigenvalues and eigenvectors of this covariance matrix are then computed. The eigenvectors, or principal components, represent the directions in the new coordinate system, and the eigenvalues correspond to the magnitude of the variance in these directions.

Mathematically, if \( X \) is the neural time series with \( n \) time points and \( p \) neurons, with a covariance matrix \( C \). The eigenvectors \( v_i \) and eigenvalues \( \lambda_i \) of \( C \) satisfy the equation:

\[ Cv_i = \lambda_i v_i \]

The neural data is then projected onto the principal components to obtain the transformed data:

\[ Z = XV \]

where \( Z \) is the matrix representing the new neural time series in the space defined by principal components, and \( V \) is a matrix of weights. Neuroscience studies using this method can then plot their data using the first two or three principal components (enabling the visualization of high-dimensional data in two or three dimensions that represent a large amount of the variance in the data), or relate the activity of these principal components to different task-related variables \cite{ebitz2021population,mante2013context,cunningham2014dimensionality}.

However, the principal components found using PCA do not separate neurons into different subsets (sub-populations). Instead, every neuron is given a continuously valued weight in the matrix \( V \) to define its contribution to each principal component. In this way, every neuron contributes in some way to every principal component. In this way, even if we develop a method for identifying significant principal components \cite{lopes2013detecting} to limit the number of components under consideration, we still have the problem of neurons belonging to multiple components.

While it is certainly possible to apply additional methods to weights of matrix \( V \) to find nonoverlapping sets of units that contribute most to each principal component, if the modular hypothesis is true, it would be useful to have a dimensionality technique that can both reduce dimensionality and identify modules out-of-the-box. Network science offers a tool to solve this problem: modularity maximization. This method uses the correlation matrix, and treats it as a network where correlation values define the strength of connections between units (e.g. a functional connectivity network). Intuitively, in the same way that PCA captures directions of shared variance, modularity maximization, by clustering together highly correlated neurons, effectively maximizes the shared variance within modules, while also separating neurons into distinct subsets. To be clear, this is a property shared by other clustering techniques (e.g. k-means clustering). 

Modularity maximization \cite{newman2004finding} is based on a straightforward principle: comparing the observed connectivity data (here defined using a correlation matrix) with what is expected by chance. This entails comparing the observed connectivity matrix \(A\), with another matrix of the same dimensions \(P\). The elements \(P_{ij}\) of matrix \(P\) represent the expected weight of the connection between nodes \(i\) and \(j\) under a null model. From these two matrices we define the modularity matrix, $B$, as:

\[B = A - P.\]

 \noindent Each element \(B_{ij}\) in the modularity matrix represents whether the observed connection between nodes \(i\) and \(j\) is stronger (\(B_{ij} > 0\)) or weaker (\(B_{ij} < 0\)) than expected under the null model. Modularity maximization then uses the modularity matrix to assess the quality or ``goodness" of a modular partition, which is a division of the network's nodes into non-overlapping modules or communities. The quality of a partition is quantified using the modularity function, \(Q\), calculated as:

\[
Q = \sum_{ij} B_{ij} \delta(\sigma_i, \sigma_j).
\]

\noindent where \(\delta\) is the Kronecker delta function, and \(\sigma_i\) and \(\sigma_j\) are the community labels of nodes \(i\) and \(j\), respectively. 

In addition to simply assessing the quality of a given partition, the variable $Q$ can be optimized outright to identify high-quality partitions of a network's nodes into modules. We used the Louvain algorithm to optimize $Q$. For a tutorial on using this method, see \citet{esfahlani2021modularity}. 

While traditionally used for binary or positively weighted networks, variants of this method have also been applied to correlation matrices (where weights can be positive or negative) \cite{rubinov2011weight}. In this context, it serves as an effective tool for dimensionality reduction and module identification. The implementation of modularity maximization that we used \cite{rubinov2011weight}  employs a degree, weight, and strength preserving null model, ensuring the identified modules are not due to these features. We also used a variant of the modularity function 
\emph{Q} that separately accounts for positive and negative edges \cite{rubinov2011weight}. For an implementation of this method, see the $\textbf{community\_louvain}$ function in the brain connectivity toolbox: \url{https://sites.google.com/site/bctnet/}.

\begin{figure*}[!t]
    \centering
    \includegraphics[width=1\textwidth]{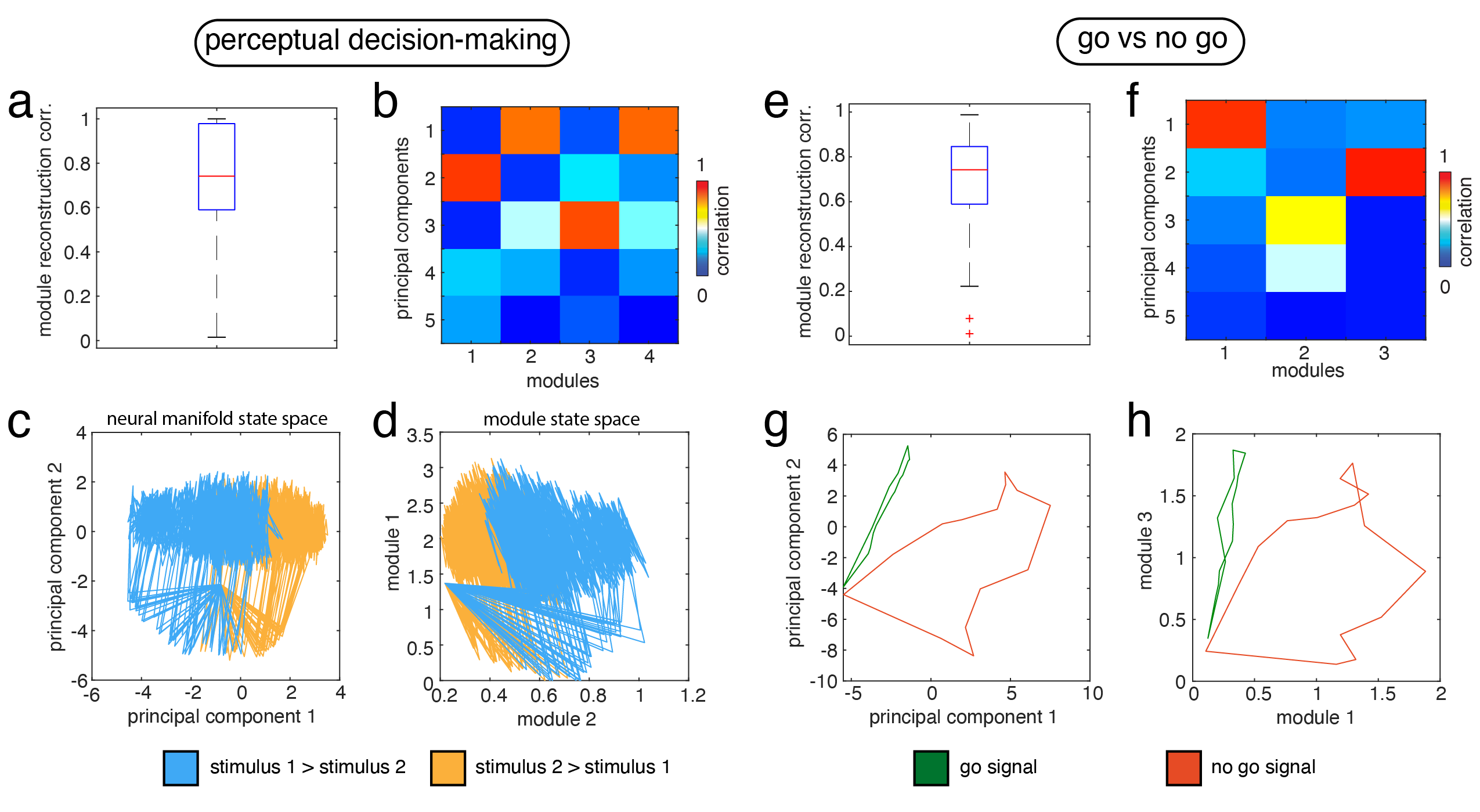}
	\caption{\textbf{Relationship between PCA and modularity maximization for dimensionality reduction.}, (\emph{a/e}) Box plot showing the correlation between original and reconstructed data when using modularity maximization as a dimensionality reduction technique with RNN activity data from a perceptual decision-making task/go \emph{vs} no go task. (\emph{b/f}) Correlation matrix showing the correlation between the activity of the first five principal components and all four modules found using modularity maximization. (\emph{c/g}) Trajectories in neural manifold state space (first two PCs) colored by correct answer for each trial. (\emph{d/h}) Trajectories in module state space colored by correct answer for each trial. } 
	\label{pca_vs_modules}
\end{figure*}

In its simplest formulation, the version of modularity maximization used in this paper tends to place neurons whose activity is positively correlated into the same module, and neurons whose activity is negatively correlated into different modules. By taking the mean activity of neurons within each module, we have reduced the dimension of the system to the number of modules. We can reconstruct the original data by mapping module activity back to each neuron (in accordance with its modular allegiance). Using artificial neural activity from an RNN trained on two different tasks (perceptual decision-making [PDM] and go \emph{vs} no go[GNG]) we find that the average correlation between reconstructed data and original data using this technique is reasonably high (Fig. \ref{pca_vs_modules}a,e; PDM mean reconstruction correlation across 2200 time points $r = 0.71$, GNG mean reconstruction correlation across 1500 time points $r = 0.69$). Additionally, because modularity maximization captures shared variance in a manner similar to PCA, the module activity captured by modularity maximization is highly correlated to the first set of principal components (Fig. \ref{pca_vs_modules}b,f; PDM mean $r = 0.89$, GNG mean $r = 0.95$). Because of this, neural trajectories in both of these spaces look highly similar (Fig. \ref{pca_vs_modules}c,d,g,h).

\section*{Results}

Throughout this paper, we explore the low-dimensional dynamics of neural activity by dividing artificial neurons into distinct modules using a community detection technique from network science known as modularity maximization. In order to assess the suitability of this method for uncovering causally important distinctions between modules, we investigate the boundaries between groups of neurons (modules) identified by this method.

The following results are organized into three sections: 1) Representation and selectivity of modules, 2) Origin of modules, and 3) Dynamics of modules. Each section asks an important question about these modular boundaries and their relevance to notable features in neural systems.

\subsection*{Representation and selectivity of modules}

Given that previous research has found principal components that are selective for different task-related variables (e.g. \cite{mante2013context}), here we investigate if modules -- being highly related to principal components -- are selective for different task-related variables. If so, this suggests that modules (as units identifying distinct subsets of neurons) also share this important feature while enabling us to lesion the system to learn more (arguably, a feature not shared with principal components).

First, we trained an RNN to solve a perceptual decision-making task \cite{molano2022neurogym}. This task presents the RNN with two stimuli drawn from normal distributions with different means. The task of the RNN is to identify the distribution with the greater mean. This requires the RNN to track previous values of both stimuli and to compare them. The task-relevant variable for this task is the cumulative difference between the means of the two input stimuli distributions (hereafter referred to as \emph{cumulative $\Delta$ inputs}).
 After a fixation period, the RNN makes its decision based on the relative activity of two output neurons. The output neuron with the greatest activity corresponds to the RNNs decision (see Fig. \ref{mod_func}a for a schematic of this task). 

After training, we used modularity maximization \cite{esfahlani2021modularity} to divide neurons into modules based on the activity of recurrent neurons across trial periods. We found four modules (Fig. \ref{mod_func}b). The mean activity for two of these modules was highly correlated with the cumulative $\Delta$ inputs variable (Fig. \ref{mod_func}e; $r = 0.78,  p < 10^{-15}$; $r = -0.86, p < 10^{-15}$; we also trained an additional 100 RNNs and tracked these values: \emph{mean} $r = 0.73$ and \emph{std} $0.25$, \emph{mean} $r = -0.74$ and \emph{std} $0.21$ respectively). Importantly, we found that these two modules were anti-correlated. Module 1 would increase its activity when stimulus 1 was estimated to have the greater cumulative mean (cumulative $\Delta$ inputs $> 0$), and module 2 would increase its activity when stimulus 2 was estimated to have the greater cumulative mean (cumulative $\Delta$ inputs $< 0$). The other two modules were active at the start and end of each fixation period, and we're directly related to the fixation input. We therefore focused our analysis on the modules that represented the cumulative $\Delta$ inputs variable. 

The above correlation-based analysis confirms that modules can be selective to task-variables, like principal components. However, unlike principal components, modules are distinct sub-populations of neurons, and as such, are naturally suggestive of various causal interventions that we can perform on the system, enabling us to test if the information, or representations, maintained about task-related variables by modules is actually used by the system (an important test for questions about representation e.g. see \cite{beer2015information,buzsaki2019brain, cao2022putting}).

 As a preliminary investigation, we found that the activity of the output neurons were correlated with the mean activity of these two modules (Fig. \ref{mod_func}d; $r = 0.77, p < 10^{-15}$; $r = 0.80, p < 10^{-15}$; across 100 trained RNNs: \emph{mean} $r = 0.79$ and \emph{std} $0.19$, \emph{mean} $r = 0.78$ and \emph{std} $0.18$ respectively). Given that the output neurons readout the decision of the RNN, this suggests that the information carried by these modules is used to make different decisions.

To verify this, we performed a lesioning analysis wherein we removed specific readout connections from modules to output neurons (See Fig. \ref{mod_func}i for a schematic). When we lesioned readout connections from module 1, we saw that task accuracy dropped to nearly 0\% for trials where stimulus 1 was the correct decision, while trials where stimulus 2 was the correct decision were not effected (Fig. \ref{mod_func}j; two-sample \emph{t}-test $p < 10^{-15}$). In contrast, lesions to readout connections from module 2 resulted in task accuracy that was high for trials where stimulus 1 was the correct decision, but destroyed accuracy for trials where stimulus 2 was the correct decision (Fig. \ref{mod_func}k; two-sample \emph{t}-test $p < 10^{-15}$). The specificity of the effects of these output lesions on different trial types suggests that the information from these modules was used by the RNN during decision-making.

\begin{figure*}[!t]
    \centering
    \includegraphics[width=1\textwidth]{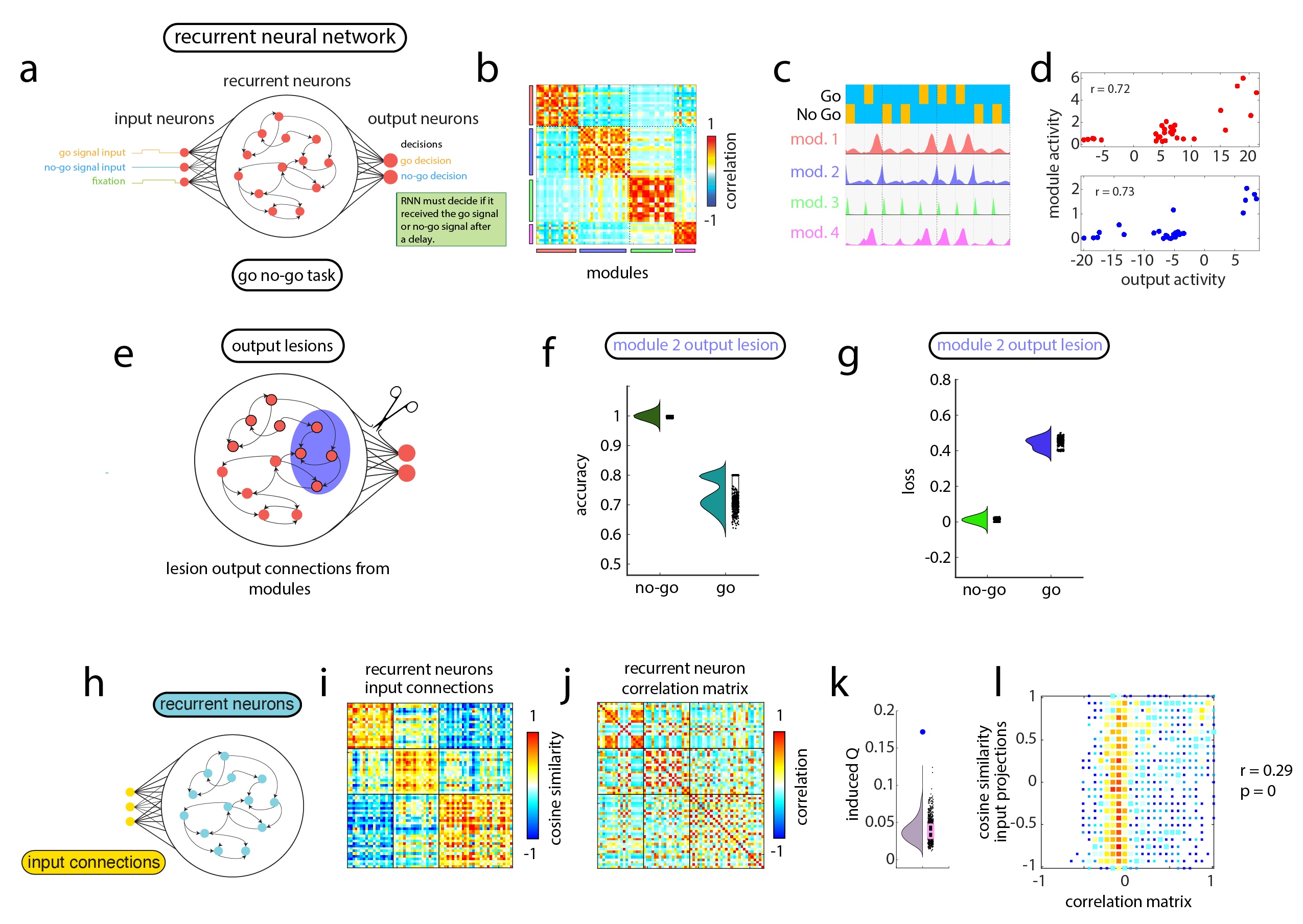}
	\caption{\textbf{Replication of RNN results with Go \emph{vs} No-Go task.}, (\emph{a}) Schematic of the architecture of the RNN used for the go \emph{vs}. no go task, and a description of the task. (\emph{b}) modules found in the activity of recurrent neurons. (\emph{c}) Mean activity within each population (color coded to correspond to the previous panel). The top of the panel indicates the signal given to the RNN (go signal, or no go signal) and the corresponding delay period. Note the increase in population one activity following the go signal. (\emph{d}) Plots showing the relationship between the mean activity in population 1(2) and output neuron 1(2). (\emph{e}) Schematic of output lesions. (\emph{f}) Boxplot showing the per trial accuracy of go and no-go conditions for the task following output lesions to population 2. (\emph{g}) Boxplot showing the per trial loss of go and no-go conditions for the task following information lesions to population 2. (\emph{h}) Schematic showing the input connections to recurrent neurons for this network. (\emph{i}) modules boundaries found using the cosine similarity of input connection weights. (\emph{j}) The correlation matrix of recurrent neurons reordered by the partition from \emph{i}. (\emph{k}) Boxplot showing the null distribution of partition quality (Q) values that we should expect by chance. This was produced by randomly permuting the partition labels from \emph{i} and applying them to \emph{j}. The real partition quality (Q) value is in blue. (\emph{l}) Plot showing the relationship between the cosine similarity of input connection weights and the correlation matrix of recurrent neurons.} 
	\label{gonogo}
\end{figure*}

We found a similar result with an RNN trained to perform a go \emph{vs} no-go task \cite{molano2022neurogym}. With this task, the RNN receives either a go signal, or a no-go signal, and after some delay must indicate which signal it received (see Fig. \ref{gonogo}a for a schematic of this task). We found four modules (Fig. \ref{gonogo}b). The mean activity with the first two modules was highly correlated to the activity of output neurons (Fig. \ref{gonogo}d; $r = 0.72$; $r = 0.73$). Finally, lesions to one of the populations significantly dropped the accuracy for trials of the task when the go signal was given, but not for trials where the no-go signal was given to the network (Fig. \ref{gonogo}e-g; two-sample \emph{t}-test $p < 10^{-15}$). 

We also trained a multi-task RNN to solve both the perceptual decision-making task and the go \emph{vs} no-go task. We analyzed artificial neural activity across 100 trials for each task, and we found that we could identify the task that the RNN was performing using only the modular partition defining the allegiance of neurons to different modules (Fig. \ref{multi-task}). This suggests that a neurons allegiance to particular modules changes in accordance with different contexts, and aligns with ideas on mixed selectivity that suggest a multi-functional role for individual neurons in different contexts \cite{fusi2016neurons,tye2024mixed,rigotti2013importance}, while also suggesting a novel relationship between mixed selectivity and a network neuroscience measure called flexibility that measures dynamic changes in module affiliation, and has been related to learning \cite{bassett2011dynamic,gu2024emergence}, memory and cognitive flexibility \cite{braun2015dynamic}, positive affect, fatigue and surprise \cite{betzel2017positive}, creativity \cite{patil2021static,li2021flexible} and intelligence \cite{barbey2018network}.

Overall, these results suggest that task-relevant information is being represented by these distinct modules and that the boundaries between these modules are informative about the function of the RNN.

In addition to our analysis of the distinct modules in RNNs, we performed a supplemental analysis to investigate modules in deep feed-forward neural networks. We found that the specific organization of neurons into modules could be used to classify input class as well as the semantic content of text-based input into a large-language model (see \textbf{Supplemental Section} \ref{sec:feed-forward}, and Figs. \ref{mnist} and \ref{transformer} for more details on our analysis).

\begin{figure*}[!t]
    \centering
    \includegraphics[width=1\textwidth]{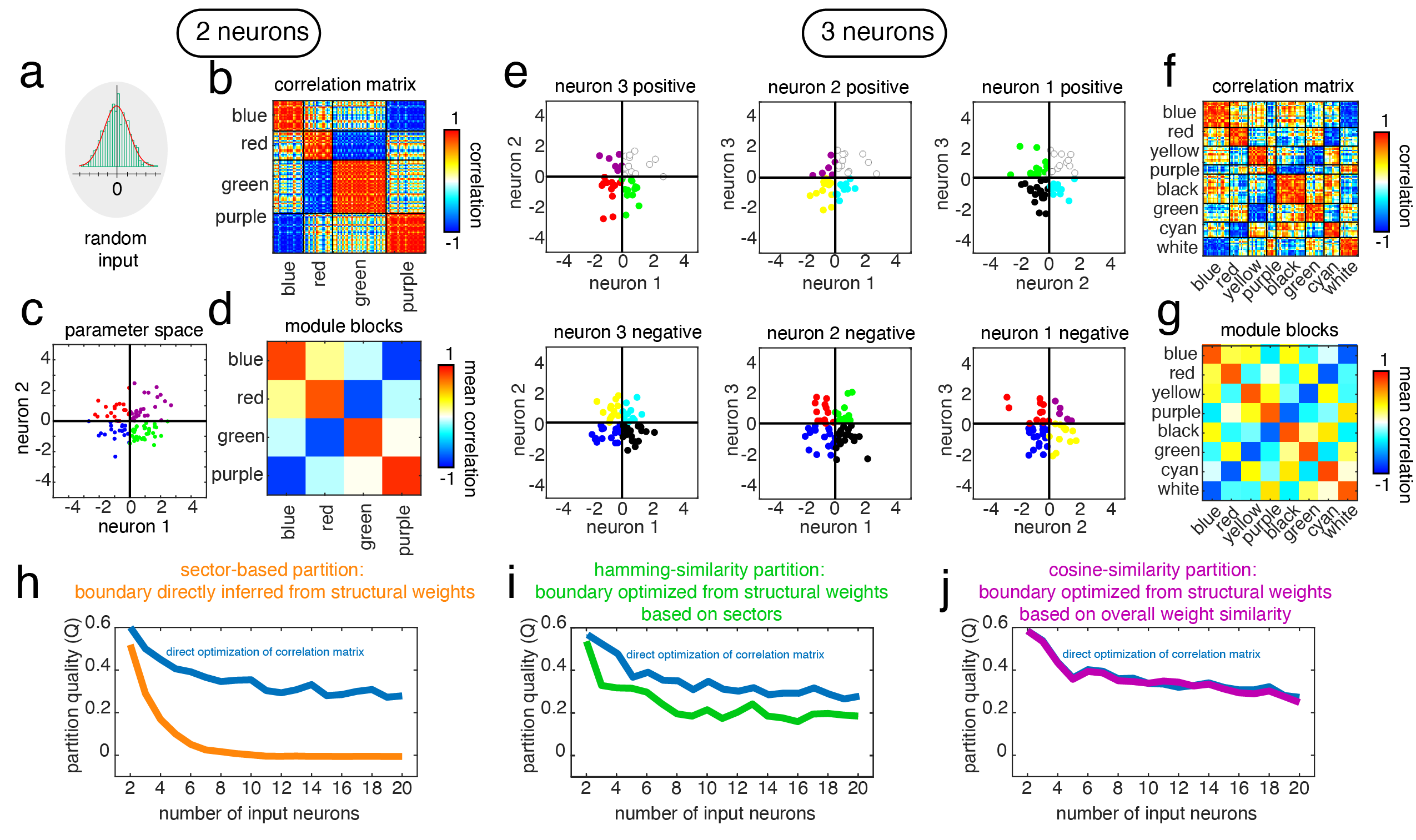}
	\caption{\textbf{Quadrants/octants and sectors determine the boundaries between modules in small N feed-forward neural networks.}, (\emph{a}) Schematic clarifying that these analyses were all performed with Gaussian distribution random input data with the same mean and variance. (\emph{b}) Correlation matrix produced by running random activity through two input neurons with random connection weights to 100 output neurons. Correlation matrix is reordered so that neurons in the same quadrant are grouped together (color labels for modules correspond to the color of dots in the next panel). (\emph{c}) Parameter space for the two input neurons connection weights to the 100 output neurons. Each point represents the set of connection weights from each input neuron to a single output neuron, and the points are colored according to the quadrant they belong to. (\emph{d}) Matrix of the mean correlation values within each population by population block of the correlation matrix in panel \emph{b}. (\emph{e}) The parameter space of a feed-forward network with 3 input neurons and 100 output neurons. For clarity, panels show all six sides of the cube representing the 3-dimensional parameter space. Dots correspond to all three input neurons connection weights to a single output neuron. Colors correspond to the octant/sector that a point falls into. (\emph{f}) Correlation matrix of output neuron activity after providing the input neurons with random activity. Matrix is reordered by the octant/sector that each output neuron belongs to, and the colors labels for the modules correspond to the colored dots in the previous panel. (\emph{g}) Matrix of the mean correlation values within each population by population block of the correlation matrix in panel \emph{f}. (\emph{h}) Plot showing the partition quality ($Q$) of the sector-based partitions of the output correlation matrix as you increase the number of input neurons. Importantly, these partitions are directly inferred from the structural connection weights and evaluated on the correlation matrix. We compare this with the partition quality values for partitions that were optimized directly on the correlation matrix. (\emph{i}) Plot showing the partition quality ($Q$) of the sector-based partitions of the output correlation matrix as you increase the number of input neurons. Importantly, these partitions are optimized based on the hamming similarity of the structural connection weights and evaluated on the correlation matrix. We compare this with the partition quality values for partitions that were optimized directly on the correlation matrix. (\emph{j}) Plot showing the partition quality ($Q$) of the cosine similarity partitions of the output correlation matrix as you increase the number of input neurons. Importantly, these partitions are optimized based on the cosine similarity of the structural connection weights and evaluated on the correlation matrix. We compare this with the partition quality values for partitions that were optimized directly on the correlation matrix. } 
	\label{IW_creation}
\end{figure*}

\subsection*{Origin of modules}

 Here, we describe the results of our investigation into the development of these modular populations. If we can find general principles that determine the boundaries between modules, then perhaps these boundaries are not merely an artifact of our methodology. By focusing on the features of the structural weight matrices defining the connection strength between every neuron, we hoped to find features that could be used to determine the boundary between the modules that emerge. This involves partially solving the specific structure-function problem of mapping structural features of connectivity onto the emergent functional properties of the network (e.g. the selectivity of neurons for different task variables).

\subsubsection*{Connection weights determine partition boundaries in feed-forward neural networks}

Where does modular neural activity come from? In large part, this is a question about how structure relates to function. The weights of the RNN implement a particular dynamical system whose lower-dimensional activity implements behavior. The question about how these weights produce lower-dimensional activity can be posed in two basic ways, depending on our assumptions about the nature of the lower-dimensional activity. If we assume neural populations are implemented in manifolds--that all neurons participate in to one degree or another-- then we ask about how the structure of neuronal connection weights constrains the behavior of neurons to covary along this subspace. In contrast, if we assume that neural populations are modular, we can ask how features of neuronal connection weights create boundaries between sets of neurons, and thus forms modules. 

Approaching this question directly for recurrent neural networks is challenging. For this reason, we aimed to simplify this problem in two important ways. First, recurrent neural network can be \emph{``unrolled''} into a feed-forward neural network with many layers \cite{sherstinsky2020fundamentals}. Using this conceptual foundation, we consider the contributions of network weights in feed-forward neural networks to the formation of module boundaries. The second simplification that we make is to consider smaller networks. For example, instead of an $M \times M$ weight matrix, we consider an $M \times N$ weight matrix where $N < M$. 

This makes the matrices that we consider here of the same type as the weight matrix defining connections from input neurons to recurrent neurons in our RNN (see Fig. \ref{mod_func}a). These simplifications made our initial search for connection weight-based rules that define modular boundaries much simpler.

In these small $N$ feed-forward neural networks, we found that the boundary between modules is determined by the sign of incoming neuronal weights. We explored this by creating connection weight-based partitions of the network into modules. More specifically, we produced output activity with such feed-forward networks by feeding the network random Gaussian input. We then computed a correlation matrix for the output activity, and used connection weight-based rules to partition the network into modules. Finally, we tested the quality of these partitions using the partition quality metric $Q$. We can understand the relative quality of this $Q$ value by comparing the partition quality ($Q$) of our connection weight-based rules with the partition quality of a partition that was directly optimized on the correlation matrix using the $Q$ metric (i.e. modularity maximization).

\begin{figure*}[!t]
    \centering
    \includegraphics[width=1\textwidth]{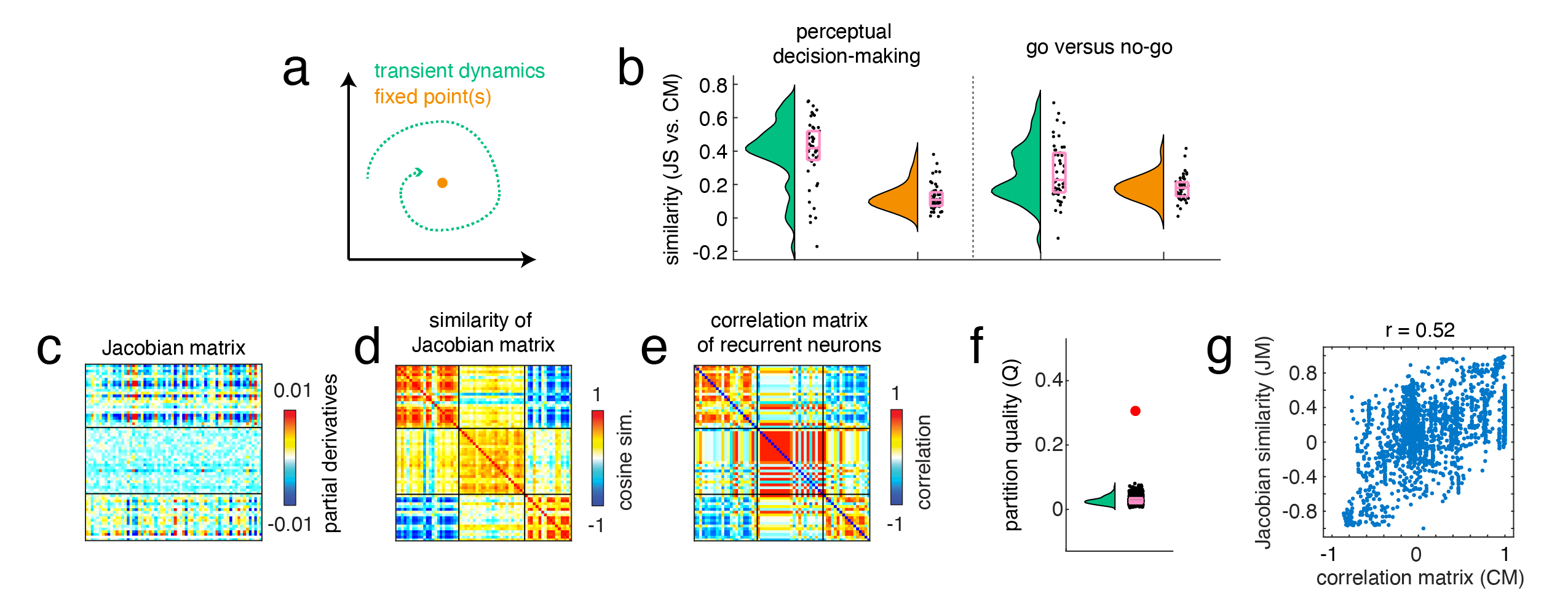}
	\caption{\textbf{Pairwise similarity of Jacobian matrix predicts modular structure.}, (\emph{a}) Schematic of showing a trajectory through state space (transient dynamics) and a fixed point. The colors of each correspond to the violin plots found in panel \emph{b} where we used a Jacobian matrix calculated from either the transient dynamics across a task trajectory (green), or near fixed points (orange). (\emph{b}) The violin plots show the correlation between the cosine similarity of rows of the Jacobian matrix and elements of the correlation matrix for recurrent neurons across 50 separately trained models on the perceptual decision-making task and the go versus no-go task. (\emph{c}) Example average Jacobian matrix across 3 task trials on the perceptual decision making task. This matrix was reordered according to the similarity of the rows in this matrix. Lines demonstrate the boundaries between modules shown in panels \emph{d} and \emph{e}. (\emph{d}) Matrix showing the pairwise cosine similarity of the Jacobian reordered according to a modular partition that was optimized on this same matrix. (\emph{e}) The correlation matrix of recurrent neurons reordered using the modular partition that was optimized on the pairwise cosine similarity of the Jacobian. (\emph{f}) Plot showing the induced partition quality $Q$ value when applying this modular partition (boundary) to the correlation matrix of recurrent neurons, compared to the null distribution when this partition was randomly permuted 1000 times. (\emph{g}) Plot showing the Jacobian similarity values plotted against the correlation values between recurrent neurons activity. } 
	\label{Jacobian}
\end{figure*}

Consider the case of two input neurons ($N = 2$). The parameter space for these neurons is two-dimensional, with one axis defining the weight of the connection from neuron 1, and the other axis defining the weight of the connection from neuron 2 (Fig. \ref{IW_creation}c). This two dimensional parameter space has four quadrants whose boundaries are defined by changes in sign (+/-). For each $M$ output neuron, the input weights from neuron 1 and neuron 2 will occupy a point in one of these quadrants. Our results suggest that the boundaries between modules are the same as these boundaries between quadrants (Fig. \ref{IW_creation}a-g).

We also demonstrate this in the case of three input neurons ($N = 3$; Fig. \ref{IW_creation}e-g), but the number of these sectors scales as $2^{N}$, such that for three neurons ($N=3$) there will be 8 sectors or octants, and for four neurons ($N=4$) there will be 16 sectors. Generally, this exponential increase in the number of sectors means that modules will become increasingly small as $N$ increases (quickly making the number of sectors/modules greater than the number of output neurons). Relatedly, as $N$ increases the quality of these sector-based boundaries decreases (Fig. \ref{IW_creation}h). We suspect this is because as $N$ increases the connection weights of each output neuron become more likely to belong to a unique sector.

We hypothesized that a process that we call ``sector collapse'' accounts for the relationship between sectors and the modules in correlation networks for larger $N$ values. Sector collapse refers to the process whereby the boundary between nearby sectors effectively disappears. To test this, we approximated the distance between sectors using hamming distance. When applied to compare two binary vectors (where $1$'s correspond to positive weights) the Hamming distance asks the number of sectors that need to be crossed to navigate between two sectors. We found that by using modularity maximization to partition the Hamming similarity of the connection weight matrix we could produce medium to high-quality modular partitions of the correlation matrix across a broad range of $N$ (Fig. \ref{IW_creation}i). This suggests that sector boundary collapse is a reasonable model for the production of modules (in correlation matrices) in weight matrices with larger $N$.

In addition to the connection weight-based rule that forms modules boundaries involving the signs (+/-) of weights, we also found an additional boundary between sub-modules involving the difference between weight values. For example, take the two-dimensional case. Consider two output neurons ($O$) whose input weights ($I1, I2$, from input neuron 1 and 2 respectively) are both positive. In parameter space, the points representing these connections are both in the same quadrant, but for $O1$ weight $I1 > I2$ whereas for $O2$ weight $I1 < I2$. Although output neurons $O1$ and $O2$ will generally be in the same module, they will now belong to separate sub-modules. In the supplemental section \textbf{Generative models of module boundaries in feed-forward neural networks}, we show this in more detail by creating generative models of modules (See \textbf{Supplemental Section}\ref{sec:generative}).

In addition to these anaylse exploring modular boundaries, we also demonstrate that the angle between weights in the $N$ dimensional space (i.e. cosine similarity) can be used to very accurately approximate the covariance structure of neuronal activity (Fig. \ref{IW_creation}j, and Fig. \ref{cosine}). This last demonstration is not a surprise given that cosine similarity is measuring the similarity of the projection of each input neurons ($N$) activity into $M$ dimensional space (where $M$ is equal to the number of output neurons).

Interestingly, when these insights about modular boundaries in feed-forward networks are applied to RNNs, they don't directly yield the correct boundaries, but they do provide useful intuitions. For example, although the cosine similarity of the weights of recurrent neurons does not do a good job of approximating the correlation matrix of recurrent neurons, the cosine similarity of another matrix does: the Jacobian matrix.

The Jacobian matrix, a matrix of partial derivatives, is typically used in dynamical systems analysis to analyze the stability of a dynamical system. The Jacobian matrix is always taken in reference to a current state. Typically, the Jacobian matrix is taken at, or near, fixed points. This is because dynamical systems typically exhibit approximately linear behavior in the vicinity of fixed points \cite{strogatz2001nonlinear}. However, we can also take the Jacobian matrix in other areas of state space (away from fixed points). For example, we can take the Jacobian at points along trajectories through the transient dynamics of the system (i.e. areas of state space that are not in the limit set). 

When we took the cosine similarity of a Jacobian matrix taken near fixed points, it was often significantly related to the correlation matrix of recurrent neuron activity (Fig. \ref{Jacobian}b; average correlation across 50 trained for perceptual decision-making (PDM) was $r = 0.13$, and go versus no-go (GNG) was $r = 0.18$; both distributions were significantly greater than zero; one-sample $t$-test $p < 10^{-14}$). Interesting, when we took the cosine similarity of the average Jacobian matrix along task trajectories, this relationship increased (Fig. \ref{Jacobian}b; average correlation across 50 trained for perceptual decision-making (PDM) was $r = 0.4$, and go versus no-go (GNG) was $r = 0.27$; both distributions were significantly greater than the distributions for the Jacobian taken near fixed points; two-sample $t$-tests (PDM) $p = 5.63 \times 10^{-15}$ and (GNG) $p = 4.72 \times 10^{-4}$). 

To investigate if the modular structure of the correlation matrix was related to the similarity of rows in the Jacobian matrix, we analyzed an example network trained on the perceptual decision-making task in more detail. First we took the average Jacobian matrix across 3 task trials (Fig. \ref{Jacobian}c) then we took the pairwise cosine similarity between all rows of the Jacobian, and ran modularity maximization to find modules in this similarity matrix (Fig. \ref{Jacobian}d). When we applied this modular partition to the correlation matrix of recurrent neurons, we found a very good fit (Fig. \ref{Jacobian}e). This partition induced a partition quality ($Q$) value of $Q = 0.31$. We created a null distribution of $Q$ values we could expect by chance by randomly permuting this partition 1000 times, and found this induced $Q$ value to be greater than chance (Fig. \ref{Jacobian}f; $p < 10^{-15}$). Overall, the values in the Jacobian row similarity matrix and the correlation matrix of recurrent neurons were highly similar ($r = 0.52$ $p < 10^{-15}$).

Here, we demonstrated that two cosine similarity matrices -- involving connection weights and the Jacobian -- could be used to approximate correlation matrices in feed-forward neural networks and RNNs respectively. In some way, these two analyses are telling us something similar. With the cosine similarity of connection weights, we are measuring the similarity of linear projections. With the cosine similarity of the Jacobian matrix, we are measuring something analogical to the similarity of linear projections, but somewhat different.  Cosine similarity of the Jacobian now involves the similarity of the effects of small local perturbations on dynamics. Both of these are in some ways measures of the similarity of influences on \emph{local} changes in activity. 

For feed-forward networks, these influences on local activity changes are simply the connection weights and inputs, but for RNNs these local changes must also account for the current state of recurrent neurons (and therefore, they must account for dynamics). When these local changes are similar between neurons, it makes sense that their activity correlates.

In a simple sense, what all of these demonstrations suggest is that these projections (whether linear, or dynamic) are closer together within modules, and more distant between modules. Perhaps these demonstrations are all ways of exploring different facets of the correlation matrix, and its relationship to connection weights in either feed-forward neural networks or recurrent neural networks. 

Taken together, these results suggest connection weights produce clusters that are related to the clusters found in the correlation matrix. In the simplest case -- with small $N$ feed-forward neural networks -- the boundaries that separate these clusters, or modules, is clear (sign-changes in parameter space). In the case of larger networks or networks with dynamics (i.e. RNNs), the direct reason for the boundary is less clear. However, by using modularity maximization to cluster connection similarity, or Jacobian row similarity, we find that the modules in correlation networks are related to these clusters.

\subsubsection*{Modular boundaries produced by input neurons influence modules of recurrent neurons}


In the RNNs we trained here, task input is delivered to recurrent neurons through feed-forward input neurons. From the previous section, we know that the sign of the input weights in this first feed-forward layer induces a boundary between modules in the next layer of neurons. However, the next layer of neurons in our network are \emph{recurrent} making the impact of this modular boundary more complicated. 

Here, we explore this question by studying how modules stabilize in random recurrent neural networks described by the following equation:

\begin{equation}
y = f(RWy + IWri)
\end{equation}

\noindent where $y$ is the current state, $RW$ are random recurrent neuron connection weights, $IW$ are random input neuron connection weights, $ri$ is random input, and $f()$ is the sigmoid activation function.

\begin{figure*}[!t]
    \centering
    \includegraphics[width=1\textwidth]{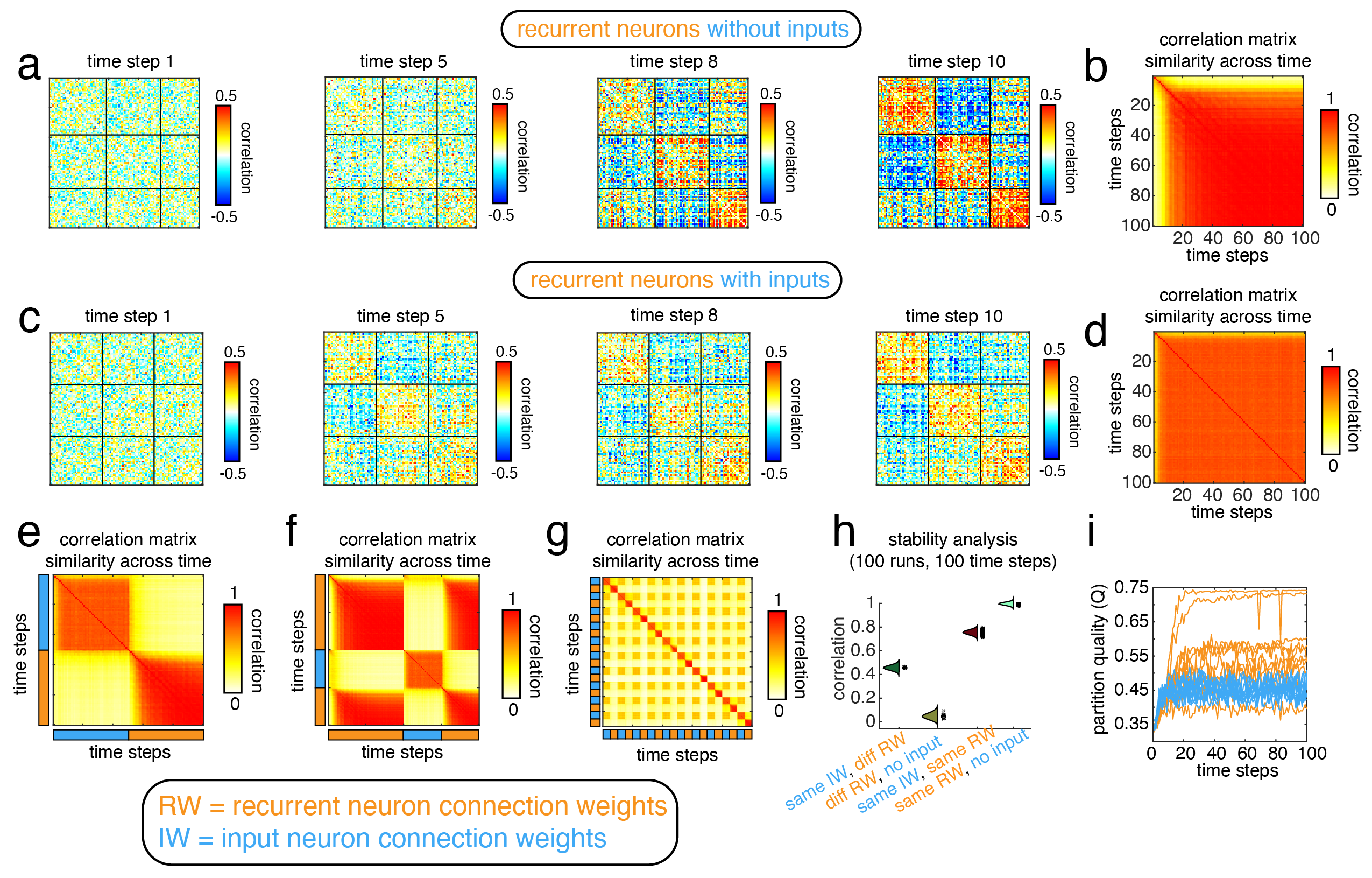}
	\caption{\textbf{Estimating the stability of covariance in large recurrent neural networks.} (\emph{a}) Here, we set the input to zero and run the RNN forward 100 steps. These panels show a sample of the correlation matrix for four of these steps. Each correlation matrix is ordered by the modules found after time step 100. (\emph{b}) Similarity matrix showing how similar the correlation matrices are across these 100 time steps. (\emph{c}) Here, we set the input to be random and run the RNN forward 100 steps. These panels show a sample of the correlation matrix for four of these steps. Each correlation matrix is ordered by the modules found after time step 100. (\emph{d}) Similarity matrix showing how similar the correlation matrices are across these 100 time steps. (\emph{e}) Here, we set the input to be random and run the RNN forward 50 steps and then removed input and ran for an additional 50 steps. This similarity matrix shows how similar the correlation matrices of activity are across all 100 steps. (\emph{f}) Here, we set the input to zero for 100 steps, then we set the input to be random and run the RNN forward 50 steps and then removed input and ran for an additional 50 steps. This similarity matrix shows how similar the correlation matrices of activity are across all 200 steps. (\emph{g}) Here, we set the input to be random and run the RNN forward 50 steps and then removed input and ran for an additional 50 steps. Then we re-randomized the recurrent weight matrix (keeping the input weight matrix the same) and reran this again (for a total of ten random $RW$ matrices). This similarity matrix shows how similar the correlation matrices of activity are across all 1000 steps. (\emph{h}) Boxplots showing the mean similarity of correlation matrices within 50 time step blocks for different conditions. (\emph{i}) The partition quality (Q) of modules across time steps in 100 models with and without input activity.    } 
	\label{rec_stability}
\end{figure*}

Our general approach was to analyze the modules that form in the presence of random input ($ri$), and to compare these modules to those that form when there is no random input (where input weights $IW$ have no effect on modules). We demonstrate that the correlation matrix in both cases stabilizes after a small number of time steps (as measured by the time-wise similarity of correlation matrices), although the correlation matrix tends to stabilize more quickly in the presence of input (see Fig\ref{rec_stability}b,d). We refer to these stable correlation matrices as ``steady-state neural populations''.

Generally, we found that the steady-state neural populations of recurrent neurons are highly different in the presence of input data (Fig. \ref{rec_stability}e,f). This suggests that the module boundaries imposed by the input neurons significantly alters the structure of populations in the recurrent neurons. But how much of the population structure in recurrent neurons can be accounted for by the boundaries from the input neurons? 

To test this, we presented many different recurrent weight matrices ($RW$) with input from the same input neurons (with the same $IW$ connection weight matrix). We found that the steady-state neural populations with the same input weight matrix ($IW$) but different recurrent weight matrix ($RW$) were highly similar (Fig. \ref{rec_stability}g,h; mean $r = 0.47$). This suggests that the modular structure in recurrent neurons carries an impression of the modules formed by the feed-forward input neurons.

To what degree can we find the the \emph{specific} module boundaries that were defined by the input neurons in the activity of the recurrent neurons? Here, we investigate this in RNNs trained on the perceptual decision-making task, and we extend this to consider a similar effect in neural connections from the thalamus to the cortex in both mice and humans.

\begin{figure*}[!t]
    \centering
    \includegraphics[width=1\textwidth]{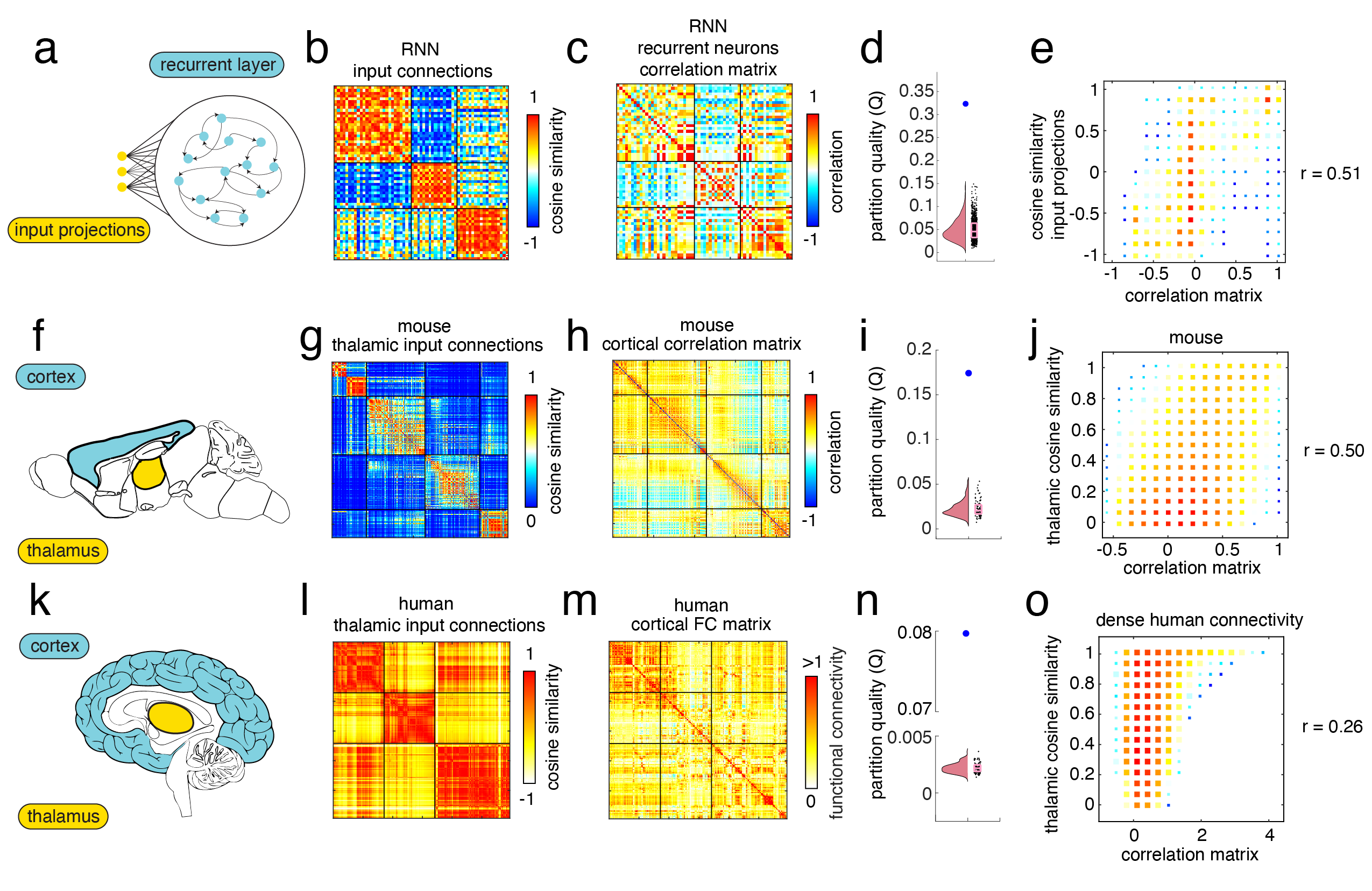}
	\caption{\textbf{Input-based moduleboundaries are reflected in recurrent neurons of RNNs and brains.}, (\emph{a}) Schematic of input connections onto the recurrent layer. (\emph{b}) Input connection similarity matrix reordered using modularity maximization. (\emph{c}) Correlation matrix of recurrent activity reordered by the partition of the . (\emph{d}) Boxplot showing the null distribution of partition quality (Q) values that we should expect by chance. This was produced by randomly permuting the partition labels from \emph{b} and applying them to \emph{c}. The real partition quality (Q) value is in blue. (\emph{e}) Plot showing the relationship between the cosine similarity of input connection weights and the correlation matrix of recurrent neurons. Dot color and size indicates the number of points that fell in this bin. (\emph{f/k}) Schematic showing the thalamus and the cortex in mice/humans. (\emph{g/l}) Thalamocortical input connection weight similarity matrix reordered using modularity maximization (for mice/humans; human matrix down-sampled for plotting).  (\emph{h/m}) Functional connectivity/correlation matrix of the cortex reordered by the partition of the thalamocortical input connection similarity matrix (for mice/humans; human matrix down-sampled for plotting). (\emph{i/n}) Boxplot showing the null distribution of partition quality (Q) values that we should expect by chance. This was produced by randomly permuting the partition labels from \emph{g/l} in a way that maintains the spatial autocorrelation in fMRI data and applying them to \emph{h/m}. The real partition quality (Q) value is in blue. (\emph{j/o}) Plot showing the relationship between the cosine similarity of input connection weights and the functional connectivity/correlation matrix of recurrent neurons. Dot color and size indicates the number of points that fell in this bin (for mice/humans).} 
	\label{empirical}
\end{figure*}

Here, we tested the quality of the modular boundaries formed by the input neuron weights when applied to the correlation matrix of recurrent neuron activity using a measure of partition quality ($Q$) \cite{newman2004finding}. We identified these modular boundaries by first taking the pairwise cosine similarity of input weights, and then using modularity maximization to find modules in this weight similarity matrix (Fig. \ref{empirical}b). By imposing this partition on the correlation matrix of recurrent neurons during task trials, we induced a modularity of $Q = 0.33$ (Fig. \ref{empirical}c). We tested if this value was greater than chance by generating a null distribution of induced $Q$ values. 

To do this, we randomly permuted the order of this partition 1000 times and calculated the $Q$ induced by imposing the permuted partition on the network for every permutation. We found that the induced $Q$ value from the input neuron partition was significantly greater than expected by our null model (Fig. \ref{empirical}d; $p < 10 ^{-15}$). Indeed, we found a strong positive relationship between the cosine similarity of the input weights and the correlation matrix of the recurrent neurons (Fig. \ref{empirical}e; $r = 0.51, p < 10 ^{-15}$) suggesting that the boundary formed by input neurons effects the modular structure of recurrent neurons.

Next, we tested if a similar phenomenon could be found in the modular structure of the cortex of mice and humans as measured by fMRI. For the rest of this analysis, we treated the thalamus as the source of input neurons projecting to the cortex. In addition to its varied roles in information propagation and modulation \cite{muller2020core,shine2021thalamus,sherman2007thalamus}, the thalamus is also the primary hub for relaying sensory information from the periphery to the cortex. As such, the thalamus might be a good analog of the input neurons found in our model. Sensory information from the eyes, ears, and body often passes through the thalamus on its path towards the cortex \cite{sherman2002role,sherman2006thalamus,sherman2007thalamus}. 

First, we tested this in mice (Fig. \ref{empirical}f). We obtained weights for the structural connections from the thalamus to the cortex using publicly available tract tracing data from the Allen Brain Institute \cite{oh2014mesoscale}. We took the cosine similarity of all in-weights from the thalamus to the left hemisphere of the cortex to create an $M \times M$ matrix, where $M$ is the number of cortical regions in the left hemisphere ($M = 2166$). We then used modularity maximization to partition this similarity matrix into modules (Fig. \ref{empirical}g). Concurrently, we analyzed functional MRI data from a separate set of lightly anaesthetized mice \cite{gutierrez2022unique}. When we applied the modular partition of the thalamocortical cosine similarity matrix onto the correlation matrix of the mouses cortical fMRI activity it induced a modularity of $Q=0.17$ (Fig. \ref{empirical}h). 

We then generated a null distribution of induced $Q$ values that we should expect given spatial autocorrelation in fMRI cortical data \cite{burt2020generative}. This involved randomly reordering nodes in a way that approximately preserved the variogram of the original data \cite{burt2020generative}. In this way, we produced 100 random partitions. We found that we could induce more modularity ($Q$) in mouse cortical FC using the connection similarity matrix partition than expected by chance (Fig. \ref{empirical}i; $p < 10 ^{-15}$). We also found a strong positive relationship between thalamocortical cosine similarity and correlation matrix of the mouse cortex (Fig. \ref{empirical}j; $r = 0.5, p < 10 ^{-15}$). After performing a similar analysis with other subcortical areas (subplate, pallidum, hypothalamus, pons, medulla, midbrain, hpf, striatum, olf, cerebellum), we found that the relationship with the thalamus was the greatest (Fig. \ref{thalamus_vs}; two-sample \emph{t}-test $p = 0.0057$). This suggests that thalamocortical connections contribute to the structure of the modular boundaries found in the mouse cortex.

Next, we tested this in humans using dense structural and functional connectivity data from the human connectome project \cite{van2013wu,seguin2022connectome,tian2021high}. 
Using the same method, we found that when we applied the modular partition of the thalamocortical cosine similarity onto the correlation matrix of the human cortex (left hemisphere;  number of nodes $M = 29696$; Fig. \ref{empirical}k-m) we could induce more modularity ($Q$) than expected by chance (Fig. \ref{empirical}i). We also found a positive relationship between thalamocortical cosine similarity and the correlation matrix of the cortex in humans (Fig. \ref{empirical}o; $r = 0.26, p < 10 ^{-15}$). 

\begin{figure*}[!t]
    \centering
    \includegraphics[width=1\textwidth]{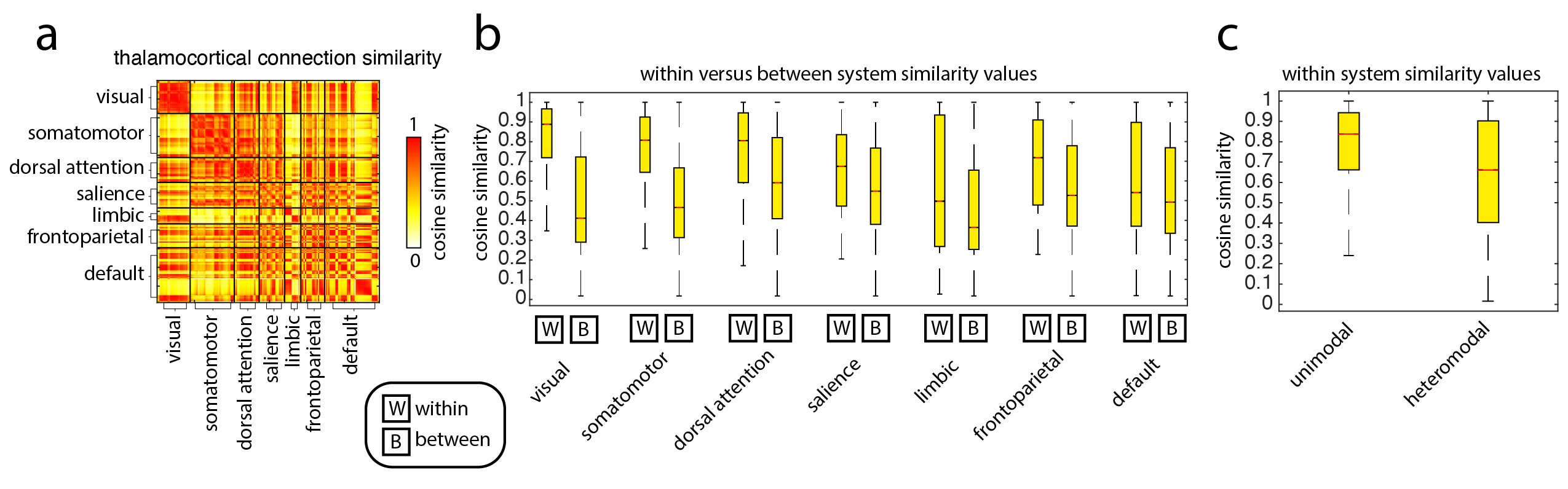}
	\caption{\textbf{Thalamocortical similarity is concentrated within seven cortical brain systems}, (\emph{a}) Thalamocortical cosine similarity matrix reordered by seven brain systems \cite{yeo2011organization,schaefer2018local}. This matrix was constructed by taking the cosine similarity of all pairs of input weights from the thalamus to the cortex with dense structural connectivity data. (\emph{b}) Boxplots comparing the thalamocortical connection similarity within each brain system versus between a given brain system and other brain systems. All brain systems have higher similarity values within the system than between systems. (\emph{c}) Focusing on the within system similarity values, we compared unimodal and heteromodal brain systems. Unimodal systems include the visual and somatomotor systems, and heteromodal contain all other systems. The unimodal brain systems have significantly higher similarity values.} 
	\label{yeo_systems}
\end{figure*}

Although this effect appears to be more diminished in the human brain, the effect is also more prominent for certain brain systems. For example, we found that the thalamocortical connection similarity values were significantly concentrated in 7 brain systems defined based on the correlational structure of resting-state fMRI across 1000 subjects \cite{schaefer2018local,yeo2011organization} (Fig. \ref{yeo_systems}a-b;  two-sample \emph{t}-test all $p < 10^{-15}$), but the brain systems for which the concentrations were highest were primary sensory systems (visual, and somatomotor; Fig. \ref{yeo_systems}c, two-sample \emph{t}-test $p < 10^{-15}$).

Taken together these results suggest that thalamocortical connections contribute to the structure of modules found in the cortex of both mice and humans. But what - if anything -  does this say about the developmental origin of modules in the cortex? We hypothesize that initial weights for the connections from the thalamus to the cortex can bias the development of coarse-grained modules. In this way, the thalamus might determine the broad placement of modules in the cortex. To support this, in a supplemental analysis we find evidence that the initial weights of input neurons influence the modules that form in recurrent neurons across different training sessions (see \textbf{Supplemental Section} \ref{sec:initial} and Fig. \ref{same_diff}).

\begin{figure*}[!t]
    \centering
    \includegraphics[width=1\textwidth]{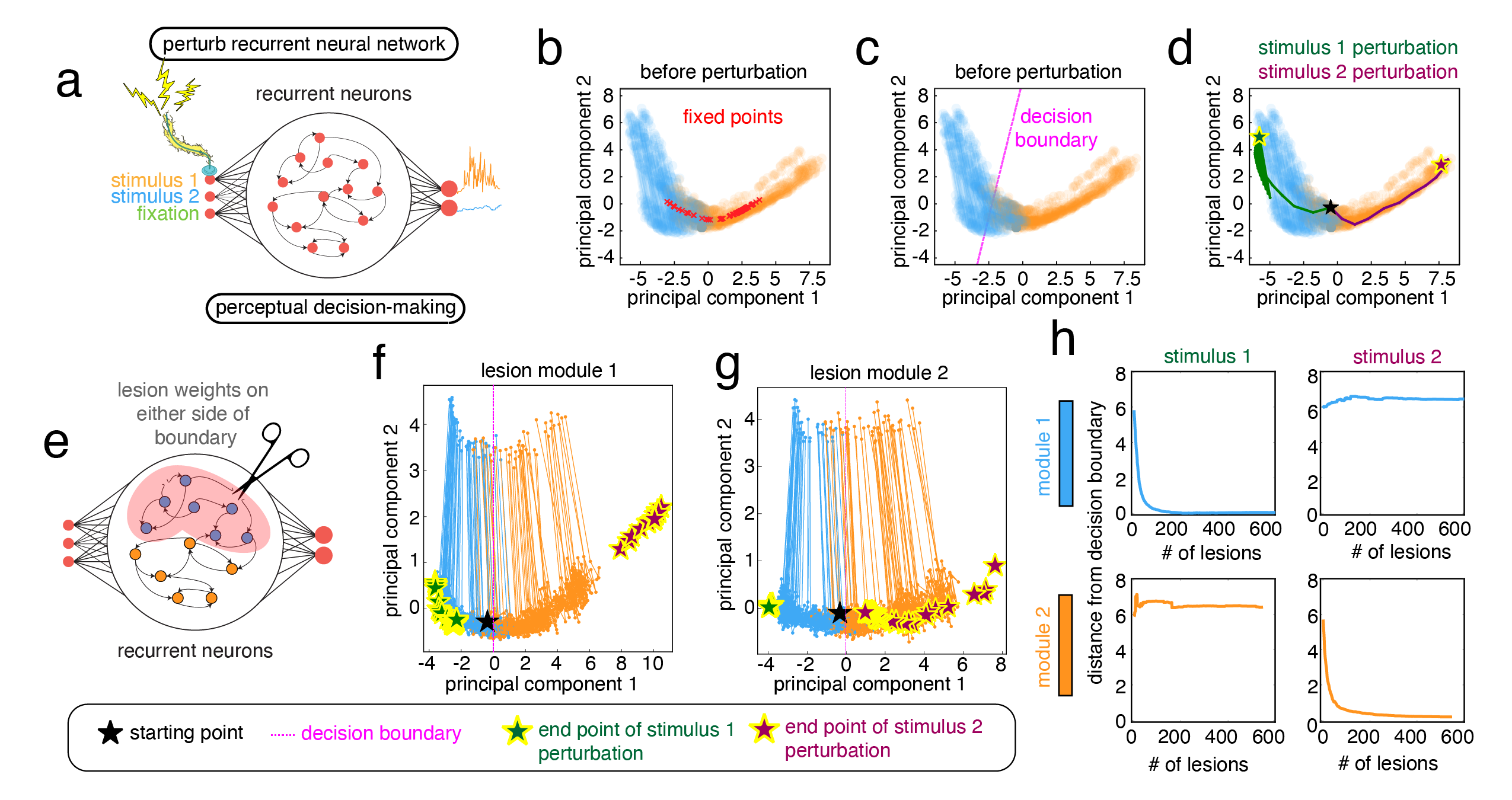}
	\caption{\textbf{Lesioning recurrent connections within moduleboundaries has circumscribed effects on dynamics.}, (\emph{a}) Schematic showing how we perturb different input neurons during our lesioning trials. (\emph{b}) Plot of the recurrent activity of a trained RNN projected into the first two principal components. In red, we plot the fixed/slow points approximated using a gradient-descent based method \cite{sussillo2013opening}. The activity trajectory is colored according to the correct decision for the trial. Note that these colors sometimes overlap given that the cumulative mean can be artificially higher for the incorrect stimulus early in the trial due to sampling variability. (\emph{c}) Same as b, but instead of plotting the fixed points, we plot the decision boundary for the network (this is a visual estimate; see main text for how the boundary was calculated). (\emph{d}) Same as b, but instead of plotting the fixed points, we plot the trajectories of two perturbation trials. In the green trial we perturbed stimulus 1. In the purple trial we perturbed stimulus 2. The state of the system starts at the black star and perturbations result in activity that stably rests at the colored stars. (\emph{e}) Schematic showing how we lesioned the weights of recurrent connection weights within moduleboundaries. (\emph{f}) Lesions to population 1 cause the end points of stimulus 1 perturbations to move closer to the decision boundary, whereas the end points for the stimulus 2 perturbation in this example move further away from the decision boundary. (\emph{g}) Showing a similar but opposite effect as j when lesioning population 2. (\emph{h}) These plots show results from our four lesioning conditions (as described in main text). When increasingly lesioning population 1, stimulus 1 perturbations move closer to the decision boundary, but stimulus 2 perturbations do not move closer to the decision boundary. The opposite is shown for increasingly lesioning population 2. These lines represent the average distance from the decision boundary across 100 trained RNNs.  } 
	\label{lesion}
\end{figure*}

\begin{figure*}[!t]
    \centering
    \includegraphics[width=1\textwidth]{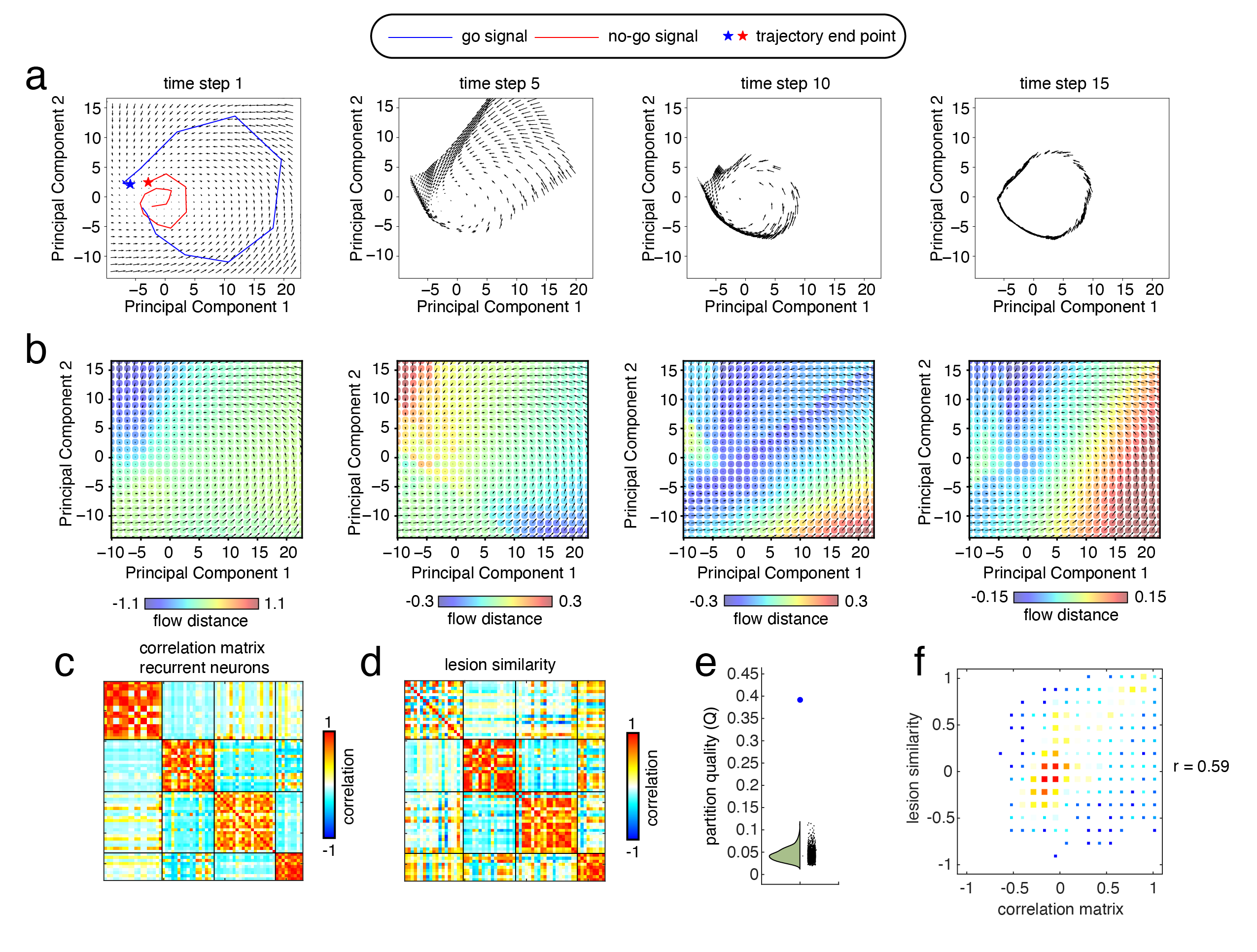}
	\caption{\textbf{Similarity of lesioning effects on flow is related to modulestructure.}, (\emph{a}) Quiver plots showing the vector fields of an RNN trained on the go vs. no-go task (projected onto the first two principal components). Recurrent activity states start in a grid of points arrayed within the total state space that is explored during task trials. Arrows show where the system will end up after a single time step (from the current time step). The arrows eventually settle onto a limit cycle. Blue and red trajectories in the first panel indicate trajectories for different task conditions. (\emph{b}) Plot of the effect of lesions on the flow shown in panel \emph{a}. A scatterplot of colored points is superimposed on a quiver plot describing the flow. The colors indicate the Euclidean distance between the non-lesioned flow after one time step and the flow after lesioning a neuron in the recurrent layer. Although plot is shown in two principal component dimensions, the distances are calculated in the original dimensions of recurrent layer activity. (\emph{c}) Correlation matrix of intact recurrent neurons. Modularity maximization was used to find modules in this matrix. This partition is also used to reorder the matrix in the next panel representing lesion similarity. (\emph{d}) We individually lesioned the weights from and to every \emph{i}-th recurrent neuron. This produced a grid of flow distance values for every neuron describing the effects of lesions on the dynamic flow. We then flattened this grid and compared the flow distances between all $i \times j$ neurons producing a matrix of similarity values telling us how similar the effects of lesions were between all neurons in the recurrent layer. This matrix was reordered by the partition of the correlation matrix found in the previous panel. (\emph{e}) Boxplot of the partition quality (Q) of the modulepartition when applied to the lesion similarity matrix. Blue dot represents real value, black dots represent a null distribution of values produced by randomly permuting this partition 1000 times. (\emph{f}) Plot showing the relationship between lesion similarity and the correlation of recurrent activity. Dot color and size indicates the number of points that fell in this bin.} 
	\label{lesion_flow}
\end{figure*}

\subsection*{Dynamics of modules}

In the previous section, we showed that the cosine similarity of rows in the Jacobian matrix could approximate the correlation matrix of recurrent neurons.  Each row of the Jacobian matrix represents the sensitivity of a particular neuron's future activity to perturbations in the activities of all other neurons in the network, and as such the Jacobian matrix plays a crucial role in providing insights about the local dynamics of the network. The relationship between the similarity of these rows and modules suggest that dynamics play a crucial role in the creation and persistence of modules. In this section, we investigate the relationship between dynamics and these modules in more detail, often using lesioning techniques. We focus on the following question: do the dynamics of recurrent neurons reflect these boundaries between modules? We investigate this question separately for the two types of boundaries that we explored in previous sections: (\emph{1}) boundaries formed by the weights of input neurons, and (\emph{2}) boundaries formed by the recurrent neurons themselves.

\subsubsection*{Dynamics reflect boundaries formed by connections from input neurons}

To test whether or not the dynamics of recurrent neurons reflect the boundaries formed by connections from input neurons, we designed a perturbation paradigm for RNNs that were trained on the perceptual decision-making task. In this perturbation paradigm, we delivered a large and continuous amount of input into one of the input neurons (corresponding to one of the stimuli (see Fig. \ref{lesion}a) for a schematic). Given such a perturbation, the RNN will infer that the artificial stimulus comes from a distribution with a much larger mean. If the dynamics responsible for maintaining and performing this inference are modular, then lesions within different modules should have relatively circumscribed effects on such inferences. We tested the hypothesis that lesions to the weights of connections between neurons within the same module would have a similar effect on this inference. 

In our calculation of the modular boundaries formed by input neurons, we ignored weights from the input neuron corresponding to the fixation (given that the fixation input neuron received tonic input of 1 across most of the trail). This resulted in two remaining input neurons. In the previous section, boundaries were formed in two ways: (1) changes in sign(+/-) of incoming connection weights ($A,B$), and (2) changes from $A > B$ to $A < B$. Here we explore the impact of both of these boundaries using a lesioning analysis. First, we explored modules whose boundaries were formed when the sign of incoming connection weights changes. Our hypothesis was that lesions to the weights connecting recurrent neurons within these modules would have specific effects on tracking the stimulus that it was associated with.

Previous work has shown that RNNs trained to perform this perceptual decision-making task set up a dynamical object referred to as a line attractor \cite{seung1996brain,mante2013context}. A line attractor is a line formed by many fixed-point attractors such that the systems long-term behavior will end up somewhere on this line in state space (Fig. \ref{lesion}b). When such a system is resting on this line attractor and is perturbed, the state of the system will return to another location along the line attractor. Because line attractors maintain the ability to hold onto incremental changes in information, they are particularly good at tracking continuous variables. In the case of this perceptual decision-making task, stimulus from each input neuron will perturb the state of the RNN in one of two general directions along the line attractor corresponding to accumulating information about the current difference between the stimulus means (this is the latent variable we explored in the first section: cumulative $\Delta$ inputs). 

Along the center of this line attractor is a decision-boundary (see Fig. \ref{lesion}c for a schematic). When the state of the system is on one side of this boundary it will make one decision, and it will make a different decision on the other side of this boundary. In this way, movement away from the decision boundary corresponds to increasing evidence for one decision and against another decision. Indeed, when we perturbed the RNN at the input for stimulus 1, the state of the system traveled away from the decision boundary and fixed itself on a leftward portion of the line attractor. In contrast, when we perturbed the RNN at the input for stimulus 2, the state of the system traveled away from the decision boundary in the other direction (for schematic see Fig. \ref{lesion}c,d). 

For our lesioning analysis, we used distance from the decision boundary as a proxy for ability of recurrent neurons to accumulate relevant evidence for each stimulus. We then trained 100 RNNs and for each RNN ran it through four lesioning conditions:

\begin{enumerate}
    \item lesion module 1 and perturb stimulus 1.
    \item lesion module 1 and perturb stimulus 2.
    \item lesion module 2 and perturb stimulus 1.
    \item lesion module 2 and perturb stimulus 2.
\end{enumerate} 

\noindent For each condition, we gradually lesioned a larger number of positively weighted connections (ordered from largest to smallest). 

\begin{figure*}[!t]
    \centering
    \includegraphics[width=1\textwidth]{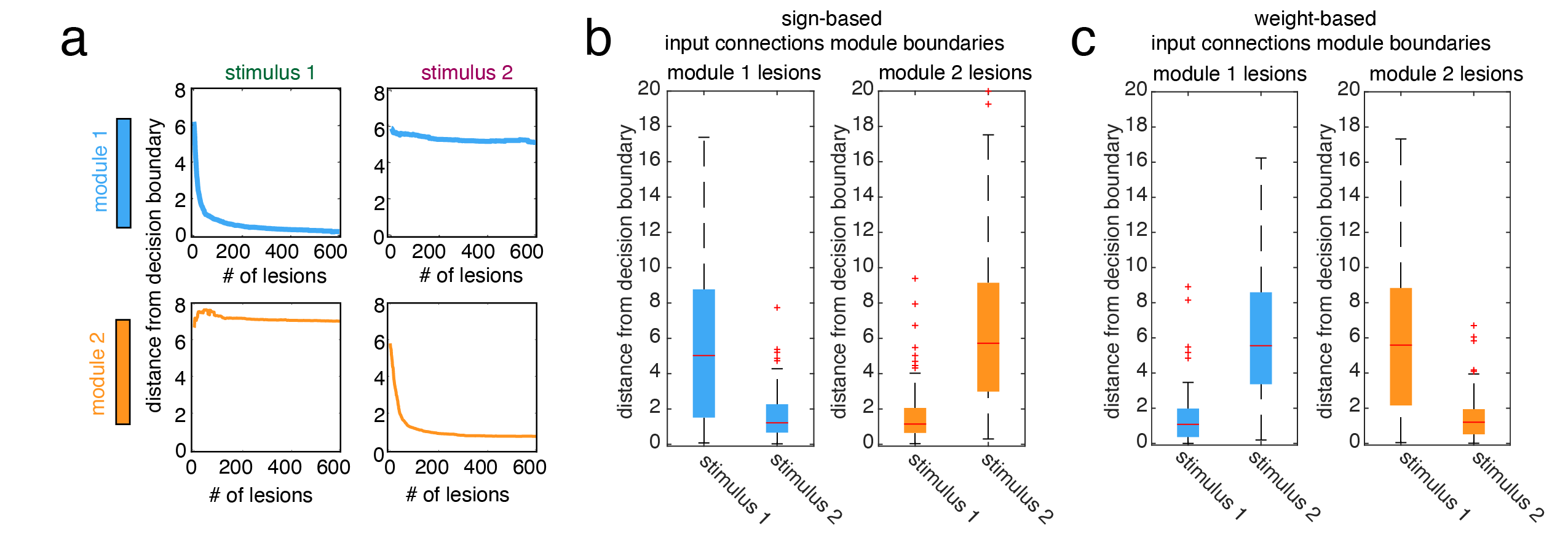}
	\caption{\textbf{Lesioning modules based on the sign/weight of the input connection weights}, (\emph{a}) These plots show results from our four lesioning conditions (as described in main text) when applied based on modules defined by the difference between input weights converging on recurrent neurons. When increasingly lesioning modules 1, stimulus 1 perturbations move closer to the decision boundary, but stimulus 2 perturbations do not move closer to the decision boundary. The opposite is shown for increasingly lesioning modules 2. These lines represent the average distance from the decision boundary across 100 trained RNNs. (\emph{b}) Boxplots for the sign-based population lesions showing the distance from the decision-boundary after 200 lesions for both stimuli perturbations. (\emph{c}) Boxplots for the weight-based population lesions showing the distance from the decision-boundary after 200 lesions for both stimuli perturbations. } 
	\label{sign_lesion}
\end{figure*}

We found that lesions to module 1 moved the state of the system closer to the decision boundary when stimulus 1 was perturbed, but had a limited effect on perturbations to stimulus 2 (sometimes causing the system to move further from the decision boundary; Fig. \ref{lesion}f). In contrast, lesions to module 2 moved the system closer to the decision boundary when stimulus 2 was perturbed, but had a limited effect on perturbations to stimulus 1 (Fig. \ref{lesion}g). We found this general feature of the specificity of lesioning to modules across 100 RNN models (Fig. \ref{lesion}h and Fig. \ref{sign_lesion}c; two-sample \emph{t}-test between stimulus 1 distances and stimulus 2 distances after 200 lesions to module1: $p < 10^{-15}$, and to module2: $p = 5.25 \times 10^{-13}$). 

We also tested this effect on modular boundaries where incoming connection weights from one neuron were greater than the other (either $A > B$ or $A < B$, where $A$ and $B$ are connection weights from two different input neurons). Using this boundary, we found the same effect (Fig. \ref{sign_lesion}a,b; two-sample \emph{t}-test between stimulus 1 distances and stimulus 2 distances after 200 lesions to module1: $p < 10^{-15}$, and to module2: $p = 2.01 \times 10^{-14}$).

Taken together, these results suggest that the boundaries between modules defined by the incoming connection weights to recurrent neurons circumscribe functionally relevant effects of lesions to recurrent neuron weights. Furthermore, this suggests that these boundaries are meaningful to the dynamics of these recurrent neurons.

\subsubsection*{Neurons within modules contribute to system dynamics in a similar manner}

In the previous section, we considered the impact of modular boundaries formed by feed-forward input neurons on the dynamics of recurrent neurons. In this section, we consider the impact of the modular boundaries present in the recurrent neurons themselves. In order to assess these boundaries, we used modularity maximization to divide neurons into modules using their activity during task trials. We then used a novel lesioning analysis to show that the modular boundaries formed by recurrent neurons circumscribe sets of neurons with similar contributions to the dynamics. 

We begin our analysis by visualizing the vector field of an RNN trained on the go \emph{vs}. no-go task (Fig. \ref{lesion_flow}a). A vector field can be used to visualize the dynamics of a system by initializing the state of that system in many different locations in state space (typically in a grid) and plotting the direction that each of these points moves after some period of time. Here, we created a grid of initial points in the two-dimensional state space defined by the first two principal components of recurrent neuronal activity. We then stepped the recurrent neurons forward for a single time step and plotted the direction and magnitude of the resulting movement through state space using a quiver plot (Fig. \ref{lesion_flow}a). After a sufficient period of time, all initial points fell onto a limit cycle (Fig. \ref{lesion_flow}b,c).

In order to quantify the effects of lesions to the weights of neurons in the recurrent layer, we analyzed how the dynamics shown in these vector fields changed following each lesion. Briefly, after lesioning a single neuron, we created a new vector field for the lesioned RNN and took the Euclidean distance between the lesioned and non-lesioned location in state space after one time step. Note that this distance was calculated in the original dimensions of the system (where $N = 100$, representing the number of recurrent neurons; we refer to this distance as \emph{flow distance}, see Fig. \ref{lesion_flow}d-f). 

After lesioning the in-going and out-going connection weights for every recurrent neuron separately, we had an array of flow distance values for each recurrent neuron. We then measured the similarity of the dynamical effects of each neuron on the vector field by taking the Pearson correlation between every $i \times j$ pair of flow distance arrays, resulting in an $N \times N$ matrix of similarity values, where $N$ is the number of recurrent neurons.

We found that this matrix of lesion similarity values was highly related to the correlation matrix of recurrent neuron activity. Not only was there a positive linear relationship between lesion similarity and FC values (Fig. \ref{lesion_flow}f, $r = 0.59, p < 10^{-15}$), but we also found that a modular partition of the correlation matrix could be used to identify modules in the lesion similarity matrix (Fig \ref{lesion_flow}d,e; $p < 10^{-15}$).

\begin{figure*}[!t]
    \centering
    \includegraphics[width=1\textwidth]{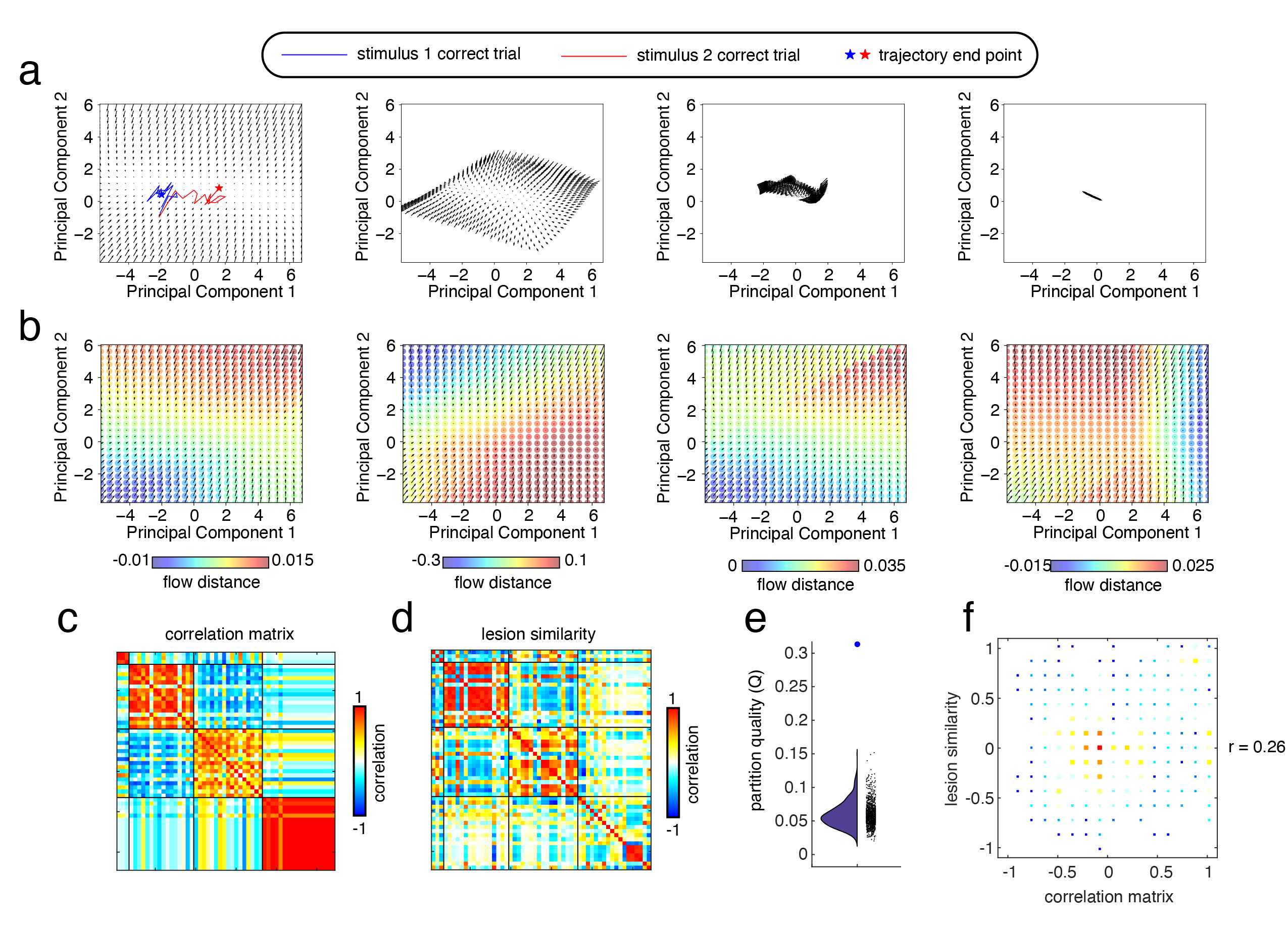}
	\caption{\textbf{Similarity of lesioning effects on flow is related to modules boundaries in RNN trained on perceptual decision making task}, (\emph{a}) Quiver plots showing the flow of the recurrent activity for an RNN trained on the perceptual decision-making task (projected onto the first two principal components). Recurrent activity states start in a grid of points arrayed within the total state space that is explored during task trials. Arrows show where the system will end up after a single time step (from the current time step). The arrows eventually settle onto a line attractor. Blue and red trajectories in the first panel indicate trajectories for different task conditions. (\emph{b}) Plot of the effect of lesions on the flow shown in panel \emph{a}. A scatterplot of colored points is superimposed on a quiver plot describing the flow. The colors indicates the difference between the non-lesioned flow and the flow after lesioning a recurrent neuron. (\emph{c}) modules boundaries of intact recurrent neurons. Modularity maximization was used to find modules in this matrix. This partition is also used to reorder the matrix in the next panel representing lesion similarity. (\emph{d}) We individually lesioned the weights from and to every \emph{i}-th neuron in the recurrent layer. This produced a grid of flow distance values for every neuron describing the effects of lesions on the dynamic flow. We then flattened this grid and compared the flow distances between all $i \times j$ neurons producing a matrix of similarity values telling us how similar the effects of lesions were between all recurrent neurons. This matrix was reordered by the partition of the correlation matrix found in the previous panel. (\emph{e}) Boxplot comparing the partition quality (Q) of the lesion similarity matrix when using the modules partition to a null model that randomly permuted this partition 1000 times. (\emph{f}) Plot showing the relationship between lesion similarity and the correlation matrix of recurrent activity. Dot color and size indicates the number of points that fell in this bin.} 
	\label{lesion_flow_perc_dec}
\end{figure*}

We replicated these results in an RNN trained on the perceptual decision-making task (Fig. \ref{lesion_flow_perc_dec}). Again we found a positive relationship between the correlation matrix of recurrent neurons and lesion similarity (Fig. \ref{lesion_flow_perc_dec}c,d,f; $r = 0.26, p < 10^{-15}$), and a significant relationship between the modules in the correlation matrix and the modules in the lesion similarity matrix (Fig. \ref{lesion_flow_perc_dec}d,e; $p < 10^{-15}$). These results suggest that modular  boundaries in recurrent neurons circumscribe sets of neurons with similar contributions to the dynamics of the system. 

\begin{figure*}[!t]
    \centering
    \includegraphics[width=1\textwidth]{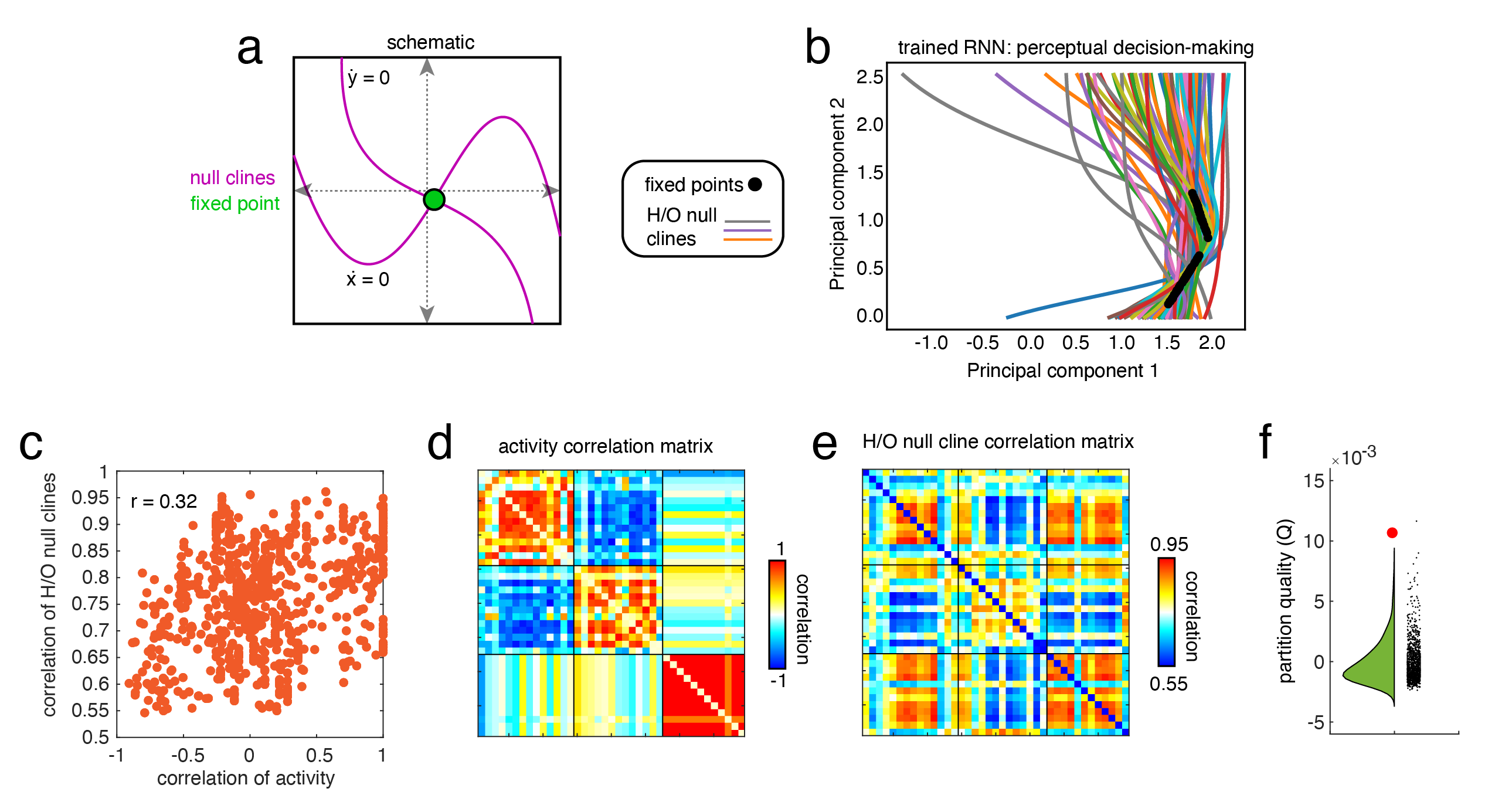}
	\caption{\textbf{Similarity of higher-order nullclines is related to modules structure.}, (\emph{a}) Schematic of the nullclines in a two-dimensional dynamical system. (\emph{b}) Approximated higher-order nullclines and fixed points in RNN trained on perceptual decision-making task (with 64 recurrent neurons). (\emph{c}) We performed a pair-wise correlation of the approximated higher-order nullclines and found that these correlation values were similar to the correlation values for the activity of recurrent neurons ($r=0.32, p < 10^{-15}$). (\emph{d}) Correlation matrix for activity of the recurrent neurons during task trials reordered into modules. (\emph{e}) Nullcline similarity matrix for all approximated higher-order nullclines, reordered by the same partition in the previous panel. (\emph{f}) Boxplot comparing a null distribution with the partition quality (Q) of the modules partition when applied to the nullcline similarity matrix (red dot). Null distribution was created by randomly permuting the order of the partition 1000 times ($p = 2.86 \times 10^{-9}$).} 
	\label{null_clines}
\end{figure*}

In order to explore this relationship further, we estimated nullclines and compared their structure with module boundaries. Nullclines are essential to the structure of the dynamics in a dynamical system. A nullcline exists where one of the state variables no longer changes its state. Where all the nullclines for a system intersect, fixed points are formed (See Fig. \ref{null_clines}a for a schematic). Deformations of these nullclines (e.g. due to parameter/weight changes) are the cause of bifurcations to system dynamics. However, as important as nullclines are to the systems dynamics, they are typically difficult to investigate in higher-dimensional systems. For this reason, we explored the use of an optimization technique to find nullclines. This technique involved minimizing the difference between the current state of neuron $i$ and the future state of neuron $i$ many times, effectively sampling from the nullcline (see \textbf{Materials and Methods}).

As we sampled the nullclines, it became apparent that they were highly complex. In order to investigate a more manageable structure while still retaining an accurate approximation of the systems dynamics, we approximated higher-order nullclines. We define higher-order nullclines as locations in state space where the derivative of all but one of the neurons is at, or close to, zero. For an RNN with 64 recurrent neurons, there are 64 of these higher order nullclines (see Fig. \ref{null_clines}b for a plot showing these approximated nullclines and the fixed points formed by their overlap).

Generally, we found that the pairwise similarity of our estimated nullclines is related to the recurrent neurons correlation matrix (Fig. \ref{null_clines}c; $r = 0.32, p < 10^{-15}$). Additionally, the nullcline similarity matrix contains modules that can be roughly mapped onto the modules in the correlation matrix (Fig. \ref{null_clines}d-f; $p = 2.86 \times 10^{-9}$).

Taken together, our results suggest that the boundaries between modules can also be used to divide the contributions of different neurons to the systems dynamics.

    \section*{Discussion}

\emph{``The fact that many complex systems have a nearly decomposable, hierarchic structure is a major facilitating factor enabling us to understand, to describe, and even to ``see'' such systems and their parts.''}

\hfill

-Herbert Simon, \emph{The Architecture of Complexity} 1962 \cite{simon1991architecture}

In his 1962 paper on the \emph{The Architecture of Complexity} Herbert Simon outlined an essential feature of complex systems: near decomposability. Not only is this feature important to the function of a complex system, its evolvability, and so on, it is also an essential factor for facilitating our understanding of the system. This emergent feature of complex systems allows them to be decomposed into a new set of elements. In other words, we can reduce the number of dimensions under consideration by coarse-graining the system into macroscale compositions of microscale components, or \emph{modules}. 

Essential in this modular description of complex systems is that this decomposability is \emph{partial}. In this way, a decomposition of the system into modules provides an approximation of the systems behavior but the modules are not entirely independent. We can imagine placing different systems on a spectrum from decomposable to non-decomposable. At the far end of decomposablity are engineered systems with parts forming components that are designed to be replacable and independent. At the far end of non-decomposability  are systems that are entirely random, where each microscale element of the system behaves independently. In the middle of this spectrum, we find nearly decomposable systems. 

For many measures of complexity, systems are often positioned along a similar spectrum between randomness and order. Measures such as algorithmic information content have been leveraged to calculate the effective complexity of a system by tracking an inverted U-shaped curve, where the most complex systems occupy the middle of this spectrum, while systems with minimal complexity—either highly ordered or highly random—are found at the extremes  \cite{gell1995quark}.

But how can we measure the level of decomposability of a system? A useful first step might involve applying dimensionality reduction techniques to explore whether or not the system can be divided into a smaller set of components. However, this approach is ultimately insufficient because it overlooks the critical aspect of \emph{removability}. The more decomposable a system is, the more the removal of one component or module should have unique and isolated effects on the overall behavior of the system. This not only reflects the functional specialization of each module but also the system's evolvability: its ability to isolate functions to enhance adaptability by allowing modules to be modified or replaced without compromising the integrity of the whole.

Removability requires that there be a physical boundary between subcomponents. This is one potential issue with using techniques like prinicipal component analysis (PCA) to identify distinct subcomponents. Because each principal component is defined by weights on each neuron, removal of any principal component is a removal of \emph{all} neurons, and therefore also \emph{all} principal components. Modularity maximization offers a solution to this problem. Not only are the features found by modularity maximization similar to those found by PCA, but it also provides a boundary between modules. With this boundary in hand, we can remove modules to calculate how decomposable the system. In doing so, we can place systems on this spectrum from complete decomposability to non-decomposability.

Using such removal techniques on recurrent neural networks (RNNs) trained on systems neuroscience tasks, we found that the modules discovered by modularity maximization are nearly decomposable. Removing readout connections from one module had unique and relatively isolated effects on the RNNs accuracy for distinct task conditions (Fig. \ref{mod_func} \& Fig \ref{gonogo}). Similarly, removal of neural connections within one module had unique and relatively isolated effects on the ability of the network to track and maintain information relevant to distinct task conditions (Fig. \ref{lesion}). Additionally, we found that removal of neurons within the same module resulted in highly similar effects on the dynamics of the RNN, whereas removal of neurons in different modules resulted in disimilar effects on dynamics (Fig. \ref{lesion_flow} \& Fig. \ref{lesion_flow_perc_dec}). Taken together, these results suggest that the modules found by modularity maximization are nearly decomposable. In other words, these results suggest that the function and contribution of neurons to dynamics are nearly circumscribed by modular boundaries and as such the effects of insults to any module are nearly isolated.

In addition to these removal/lesion based analyses, we also conducted a number of additional analyses to explore the origin of the boundaries between modules and their relationship to system dynamics. Each of these analyses identifies a useful vantage point from which to explore the clustered structure of the correlation matrix (which is identified as ``modules'' by modularity maximization). In this way, these analyses operate like a kind of prism. A prism takes white light and separates it into its different component wavelengths (colors) through a process called dispersion. Here, we disperse the correlation matrix of recurrent neurons into component parts: partial derivatives, nullclines, feed-forward input projections, and so on.

\begin{figure*}[!t]
    \centering
    \includegraphics[width=1\textwidth]{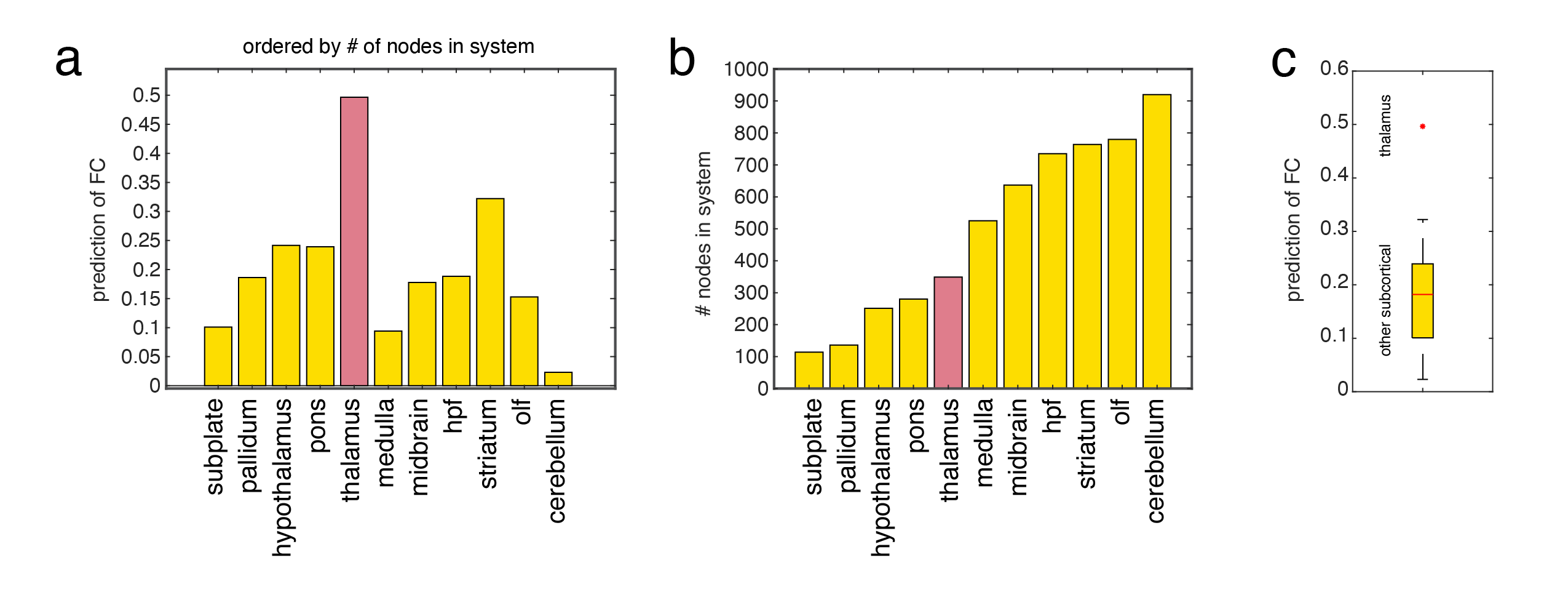}
	\caption{\textbf{Thalamocortical cosine similarity is a better predictor of the cortical correlation matrix than the cosine similarity of any other subcortical system.}, (\emph{a}) Bar plot of correlation values for cosine similarity of different subcortical systems with cortical functional connectivity/correlation matrix. This plot is ordered by the number of nodes in the system. Thalamus has the highest correlation with FC even though it has less nodes than most other systems. (\emph{b}) Bar plot showing the number of nodes in each subcortical system. (\emph{c}) Boxplot comparing the correlation values (connection similarity vs. cortical correlation matrix) for the thalamus and other subcortical structures.} 
	\label{thalamus_vs}
\end{figure*}

We find that the weights of connections from input neurons influences correlation structure not only during task performance (Fig. \ref{input_projections} \& Fig. \ref{rec_stability}), but also across training (Fig. \ref{same_diff}). We also see a similar relationship in real brains from humans and mice, wherein the input weights from the thalamus to the cortex inform us about the modules that form in the cortex (Fig. \ref{empirical} \& Fig. \ref{yeo_systems} \& Fig. \ref{thalamus_vs}). In addition, we find that the rows of the Jacobian matrix (partial derivatives for each neuron) clusters to form highly similar modules to the modules found in the correlation matrix (Fig. \ref{Jacobian}), suggesting that the modules in the correlation matrix might emerge from this clustering in the dynamics. This is further evidenced by our analysis of the higher-order nullclines for this RNN which also form clusters that are highly similar to the clusters found in the correlation matrix (Fig. \ref{null_clines}).

\begin{figure*}[!t]
    \centering
    \includegraphics[width=1\textwidth]{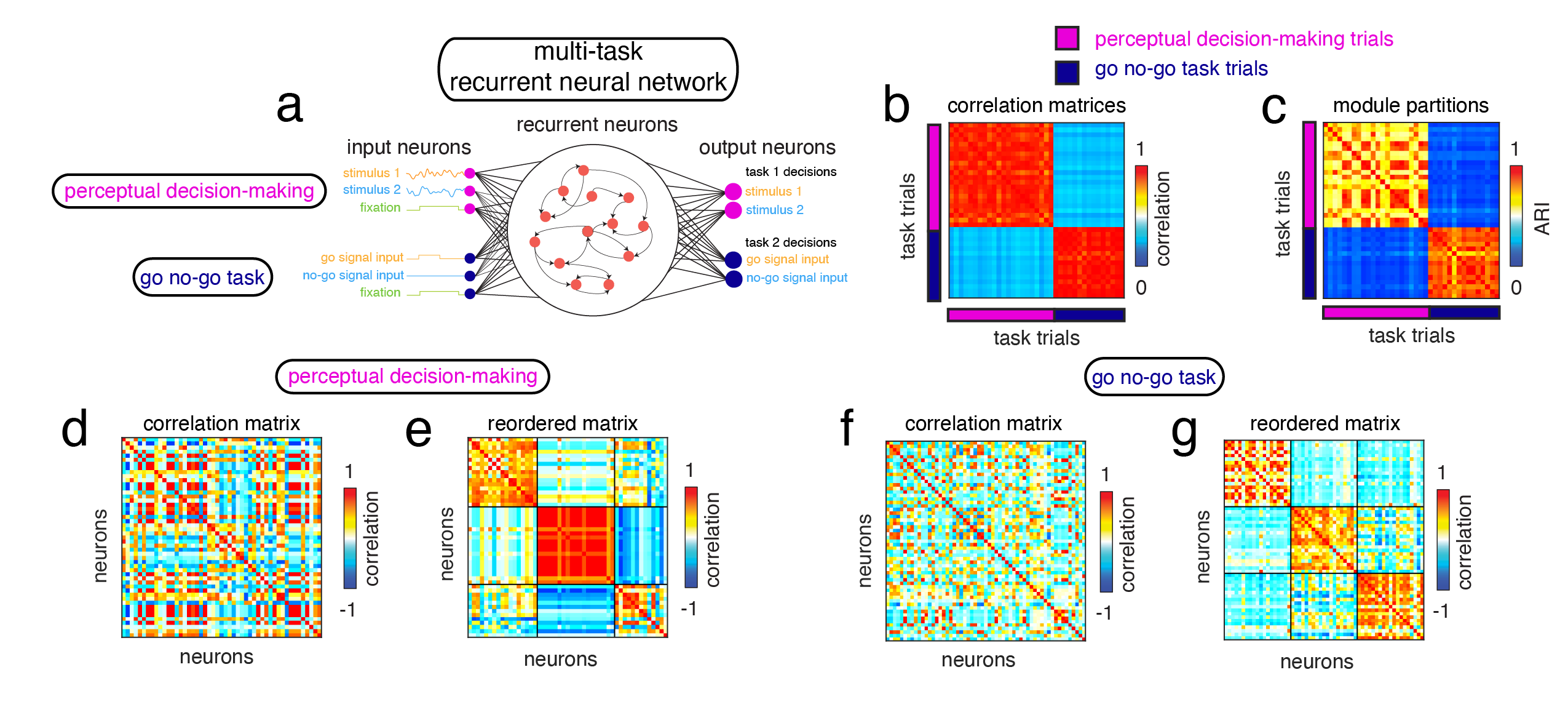}
	\caption{\textbf{Task trials in multi-task networks can be identified by modular structure.}, (\emph{a}) Schematic of the architecture of the multi-task RNN trained on the perceptual decision-making task and the go \emph{vs}. no go task. In order to investigate how the modular boundaries shift in multi-task RNNs when the RNN changes tasks, we computed a correlation matrix for each task trial. We found that correlation matrices for task trials of the same task (different trial) were more similar than different tasks (\emph{b}). We also found that modular partitions for the same task (different trial) were more similar than different tasks (\emph{c}). (\emph{d}) Correlation matrix for a perceptual decision-making task trial (standard ordering of neurons). (\emph{e}) Reordered correlation matrix for a perceptual decision-making task trial (partition from modularity maximization).  (\emph{f}) Correlation matrix for a go \emph{vs} no go task trial (standard ordering of neurons). (\emph{g}) Reordered correlation matrix for a go \emph{vs} no go task trial (partition from modularity maximization).   } 
	\label{multi-task}
\end{figure*}

Finally, we also found modular partitions (i.e. labels identifying which module a neuron belongs to) carried rich information about task structure in RNNs, simple feed-forward neural networks, and even more advanced transformer-based large language models (LLMs). We showed that the similarity of modular partitions could be used to identify which task a multi-task RNN was performing (Fig. \ref{multi-task}), which hand-written digit a feed-forward neural network was seeing (Fig. \ref{mnist}) and even identify semantic content similarity in an LLM (Fig. \ref{transformer}).

Our work adds to a rich history of literature exploring modularity in model systems. As we have shown, network science offers a compelling set of tools with which to explore modular structure by operationalizing interactions between units as the nodes and connections in a network. Modules, or communities, are groups of nodes that are more densely connected to one another than to nodes outside of the module \cite{newman2012communities}. Importantly, these network connections can be defined in a myriad of ways, leading to different explorations of modularity. 

In neuroscience, these connections are often defined as physical connections between units: whether as synaptic connections between neurons, or white matter connections between brain regions. Such networks would be considered \emph{structural} brain networks, and their modules would be structural modules. In some past theoretical/computational work, structural modules \emph{emerge} while training networks to perform tasks with a particular structure. For example, in 2005 Kashtan and Alon found that these networks would naturally produce structural modules when evolved to produce a series of interrelated \emph{modular goals}. Modular goals are goals which are themselves composed of subgoals, and these modular goals were \emph{interrelated} when the subgoals that composed each of them came from the same common set. \cite{kashtan2005spontaneous}. Similarly in 2010, Espinosa-Soto et al also found a relationship between structural modularity and the existence of multiple goals in the evolution of gene regulatory networks \cite{espinosa2010specialization}.

Whereas these papers showed that structural modularity emerged with specific task structure, Clune et al showed that you can get structural modularity for free from connection costs \cite{clune2013evolutionary}. That said, Clune et al explicitly state that they do not believe this means structural modularity is a spandrel \cite{gould2020spandrels}. Instead, they see this process of getting structural modularity for free as a bootstrapping process whereby the modularity that emerges naturally from connection costs can be reused in order for the system to become more adaptable. For example, their future work found that this ``modularity for free'' assists the network in remembering old skills while acquiring new ones \cite{ellefsen2015neural}. More recently, in 2023, Achterberg et al similarly found that recurrent neural networks trained with connection costs (while optimizing within-network signal transmission) naturally produce structural modules \cite{achterberg2023spatially}.

In contrast, other theoretical/computational work has explored the features of networks that were \emph{forced} to have structural modules. For example, in 2019 Rodriguez et al altered the structural modularity of a recurrent neural network manually in order to assess what level of modularity is optimal for a given memory task \cite{rodriguez2019optimal}. In addition to finding that modularity increased the memory capacity of the network, the authors also found that modularity increased the number of attractors available in the network. Other work found that structural modularity enabled the separation of dynamic time-scales \cite{pan2009modularity}. In 2021, Kleinman et al trained multi-area recurrent neural networks where each area is effectively a structural module, and found that inter-modular communication assisted with a kind of information filtration \cite{kleinman2021mechanistic}. 

Whereas Rodriguez et al and Kleinman et al solely enforced modular structure, in 2021 Suarez et al used empirically-derived connectomes as the connectivity structure in recurrent neural networks, and these networks naturally display structural modularity, alongside other important network features (e.g. efficient path structure, heavy-tailed degree distributions). Interestingly, these empirical networks also produce modules in the correlation matrix describing their activity (i.e. functional connectivity modules; see \cite{suarez2020linking} for review). Suarez et al found that these empirically defined functional modules (intrinsic brain networks, see \cite{yeo2011organization}) are naturally the best read-out areas for memory dynamics \cite{suarez2021learning}. 

Less work has been done exploring the modular structure of activity-based networks in model systems like RNNs. Two important recent papers find that you can cluster the variance of neural activity in multi-task RNNs \cite{yang2019task} and that these clusters correspond to task general dynamical motifs that are reused in similar tasks \cite{driscoll2024flexible}. Whereas these papers find clusters in the variance of neural activity in RNNs, we build on this work by exploring clusters/modules in covariance based networks (correlation networks). Importantly, clustering covariance means that the modules we find can also be construed and explored as a means of dimensionality reduction (Fig. \ref{pca_vs_modules}).

Finally, our work suggests that one potentially important avenue for research into neural network interpretability will be the use of tools to investigate the near decomposability of these systems. As Herbert Simon suggested, it is this feature of complex systems which enables us to understand them. When physicist Richard Feynman died in 1988, his blackboard at Caltech held a now-famous saying: "What I cannot create, I do not understand." This quote has guided many scientists and engineers with a sobering and difficult goal post: create systems that can generate the behavior of interest. However, with the advent of modern neural network models (including RNNs, and large-language models), we can create systems that behave in an intelligent manner, and nonetheless we do not fully understand their behavior. Creation, then, might be necessary for understanding, but it is not sufficient. Perhaps future neural network interpretability research could benefit from focusing on the ways in which these modern neural network models are nearly decomposable, and one such method is modularity maximization.

	\section*{Materials and Methods}

    \subsection*{Recurrent Neural Network}

We used a specific type of recurrent neural network (RNN) referred to as continuous time recurrent neural networks (CTRNNs). CTRNNs are defined by the following equation \cite{nn-brain,yang2019task,yang2020artificial,beer2006parameter}:

\[\tau \frac{d\mathbf{r}}{dt} = -\mathbf{R}(t) + f(W_{rec} \mathbf{R}(t) + W_{in} \mathbf{Input}(t) + \mathbf{b}).\]

Here, $\tau$ is a time constant (set to 100), \emph{\bf{R}} is the recurrent state, $W_{rec}$ are the weights of the recurrent layer, $W_{in}$ are the weights of the input layer, and $\mathbf{b}$ is a bias. For all implementations of RNNs in this project we chose CTRNNs with a recurrent layer size of  $M = 100$. Finally, a linear transformation of the recurrent state \emph{\bf{R}} is used to create output activity:

\[
\mathbf{O}(t) = g(W_{out} \mathbf{R}(t) + \mathbf{b}_{out}).
\]

For more information on the design and implementation of this CTRNN see the github repository \url{https://github.com/gyyang/nn-brain/blob/master/RNN_tutorial.ipynb} \cite{nn-brain,yang2020artificial}. For more information on the dynamics and parameter space of CTRNNs generally, see \cite{beer2021global,beer2006parameter,beer1995dynamics}.

    \subsection*{System neuroscience tasks for recurrent neural networks}

We used a machine learning toolbox called neurogym: (\url{https://github.com/neurogym}). The toolbox is built upon the popular machine learning framework \emph{gym} from OpenAI and provides a variety of systems neuroscience tasks that have been designed for easy presentation to recurrent neural networks (see \cite{yang2019task}). We chose two tasks from this toolbox to focus on: 1) the perceptual decision-making task, and 2) the go vs no-go task.

The perceptual decision-making task implements a simplified version of a random-dot motion task wherein the subject is presented with randomly moving dots that have coherent motion in some direction \cite{britten1992analysis,mante2013context}. The task of the subject is to indicate the direction of coherent motion during some trial period. In the neurogym version of this task, the trial period is determined by a fixation input to the RNN. Two stimuli are presented to the RNN during this fixation period and the task is to indicate which of the two stimuli has the larger mean value. Each of these stimuli comes from a Gaussian distribution with different means (equal standard deviation). The means of this distribution represent coherent motion, and the random distribution of values around the mean represent the random motion.

The go \emph{vs} no-go task also has a trial structure that is determined by a fixation input. At the beginning of the fixation period, the RNN is presented with one of two signals: a go signal or a no-go signal. These two signals are represented by two different input neurons. Following presentation of this signal, there is a delay period wherein the RNN only receives the fixation signal. The delay period ends when the fixation signal is no longer present, and the RNN must now determine which of the two input signals it received before the delay period (the go signal, or the no-go signal).

RNNs were trained on both of these tasks in PyTorch, an open-source machine learning library for the implementation and training of machine learning models using backpropagation \cite{paszke2019pytorch}. Pytorch simplifies the training of such models by automatically tracking the gradient of all computations in the forward pass of a neural network, and storing them in a computational graph that can later be used to backpropagate error.

     \subsection*{Output lesions}

 In our \emph{\bf{Results}} section on the \emph{Representation and selectivity} of modules in recurrent neural networks, we use a lesioning method that we refer to as ``output lesions" to test if the RNN is using information from its modules to perform the task. These output lesions are lesions to the weights of the connections from recurrent neurons to output neurons. The output lesions were performed in the following way:

We implemented the modularity maximization method optimized using the Louvain algorithm to partition the correlation matrix of the recurrent activity into $modules$. We then use these modules to lesion the connections from a given module to the output neurons:

\[ \text{for selected module } m, \in modules, \, W_{out}[:, m] = 0. \]

\noindent here, $W_{out}$ be the weight matrix for connections from the recurrent layer to the output layer

    \subsection*{Generative models}

In the \emph{\bf{Supplemental Section}} on Generative models of moduleboundaries (\textbf{Section} \ref{sec:generative}), we developed three generative models of moduleboundaries. Each of these generative models involved creating a new input weight matrix, and the parameters of the models allowed us to explore how partition quality, or modularity ($Q$) changed as we varied parameters in the model.

The first generative model was the sign-based model. This model creates an input weight matrix $W$ of size $N \times M$ out of two values ($+1, -1$). That is:

\[W_{ij} = 
\begin{cases} 
-1 & \text{if } i \in S \\
1 & \text{otherwise}
\end{cases}
\]

\[
S = \{i_1, i_2, \ldots, i_k\}.
\]

\[k = \left\lceil N \times p \right\rceil.\]

\noindent where $S$ is a random subset of row indices up to $k$, and $k$ is defined based on the percentage $p$ of negative weights to be added to the weight matrix $W$. All columns of this matrix $M$ are given the same value.

The second generative model was the difference-based generative model. This model creates an input weight matrix $W$ of size $N \times 2$ by randomly assigning one columns value from each row  to $\alpha$ and assigning the other columns value to $\alpha + \gamma$:

\[W_{i1}, W_{i2} = 
\begin{cases} 
(\alpha_i, \alpha_i + \gamma) & \text{with probability } 0.5 \\
(\alpha_i + \gamma, \alpha_i) & \text{with probability } 0.5
\end{cases}
\]

\[
\alpha_i = \left| x_i \right|, \text{ where } x_i \sim \mathcal{N}(0, 1).
\]

\noindent $\alpha$ is the absolute value of a random variable drawn from a Gaussian distribution with a mean of zero and a variance of 1. 

Our third and final generative model included both sign and difference elements and is likewise refered to as the sign-difference generative model. The only difference between the previous model and this model is the definition of $\alpha$.

\[
\alpha_i \sim \mathcal{N}(0, 1).
\]

\noindent That is, whereas the previous model took the absolute value of the randomly drawn variable, this model does not, allowing approximately half of these values to be negative.

    \subsection*{Fixed point approximation}

In our \emph{\bf{Results}} section on \emph{Dynamics}, we illustrate that the perceptual decision-making task results in an approximate line attractor (Fig. \ref{lesion}b). In order to approximate the attractors that form this line, we used a simple gradient based method. For the implementation of the fixed-point approximation method that we used (also showing an approximate line attractor with the perceptual decision-making task) see: \url{https://github.com/gyyang/nn-brain/blob/master/RNN%2BDynamicalSystemAnalysis.ipynb}.

Briefly, this process involves optimizing the hidden/recurrent activity of the RNN such that the mean squared error (MSE) between that activity and the hidden/recurrent activity one step forward in time is minimized. Attractors occur where the derivative of a system is equal to zero. Intuitively then, if the difference between a current state and the next state is brought to zero, the system is in an attractor state. Alternatively, the loss could be very low but not zero, a case that is sometimes referred to as a slow-point \cite{sussillo2013opening}. This optimization is implemented by randomly initializing the recurrent state of the RNN in many different states, and then using backpropagation to minimize the MSE between current activity and the future activity.

    \subsection*{Nullcline approximation}

Building on the fixed point approximation method described in the previous section, we built a simple method to approximate the nullclines/manifolds of large dimension RNNs. Nullclines exist where the derivative of one state variable is equal to zero. Briefly, we used gradient descent to approximate this by minimizing the difference between a random current state of neuron $i$ and the future state of neuron $i$. This resulted in a $J \times M$ matrix describing the state of all $M$ neurons sampled along $J$ points in the manifold. 

In the paper, we used this analysis to approximate what we called ``higher order nullclines''. These are manifolds along which the derivative of all states except for state $i$ are equal to zero. For this reason, they approximate regions in state space where all nullclines are intersecting except for the nullcline associated with state $i$.

    \subsection*{modulelesions}

In our \emph{\bf{Results}} section on \emph{Dynamics}, we explored the effects of lesions to the weights of the recurrent neurons of RNNs trained on the perceptual decision-making tasks. We used a sign-based partitions or difference-based partitions of the input weights to define the boundaries between modules (Fig. \ref{input_projections}d and i respectively). Briefly, sign-based partitions were determined by taking the sign of weights from one of the input neurons. Neurons receiving positive weights were in one neural population. Neurons receiving negative weights were in another. Difference-based partitions were determined by comparing the weights of the two input neurons onto each recurrent neuron. If  $input weight 1 > input weight 2$ then the recurrent neuron was placed in module1, and if $input weight 1 < input weight 2$ then the recurrent neuron was placed in module2.

Then, we tested the effects of lesions to RNNs that were``perturbed" with a large continual input value of 4. More specifically, we lesioned positively weighted connections in the recurrent layer that were within the boundaries defined by these modules:

\[
W'_{ij} =
\begin{cases}
0 & \text{if } i, j \in M_k \text{ and } W_{ij} = \max\limits_{a, b \in M_k}(W_{ab}),\\
W_{ij} & \text{otherwise.}
\end{cases}
\]

Here, $W$ is the weight matrix of the recurrent layer, $W'$ is the modified weight matrix after the lesioning, $M_k$ is a module in the network, and $i$ and $j$ are neurons in the recurrent layer. If neurons $i$ and $j$ belong to the same module $M_k$ and the weight $W_{ij}$ is the maximum weight among all the connections within the module, then the weight is set to zero. Otherwise, the weight remains the same. We iterated this equation such that $W'$ became $W$ on future iterations. In this way, we lesioned an increasing number positively weighted connections within the same moduleand on each iteration we tested the distance of the perturbed activity from the decision boundary. Given that these RNNs were trained on cross entropy loss, the output neuron with the greatest activity corresponded to the RNNs ``choice". For this reason, we defined the decision boundary as the difference between the output activity of the two output neurons that correspond to decision 1 and decision 2.  

    \subsection*{Lesion similarity}

In our \emph{\bf{Results}} section on \emph{Dynamics}, we explored the effects of lesions to recurrent neurons of the RNN by quantifying how they changed the vector fields of the system. First, we defined the vector field of the RNN by performing principal component analysis (PCA) on the activity of recurrent neurons across many task trials. Then, we defined a grid of $25 \times 25$ points (for a total of 625) in the 2-dimensional space defined by the first two principal components. These points are spread equally so that they span the entire activity space explored during task trials. Then, we use each of these points as an initial recurrent state for the RNN (projecting it back into the original activity space) and step it forward in time by one time step. This results in a new set of 625 points corresponding to the location of the recurrent state after one time step from each point on the original grid. We stored these points as the original points $Orig$ (importantly, we stored each of these points in the original $M$-dimensional activity space).

For the lesioning analysis, we would lesion a neuron and then use the same grid of initial recurrent states to initiate the system and run it for a single time step. This resulted in a new set of forward stepped points $L$ for the lesioned model. We created $M$ separate lesioned models where $M$ is the number of neurons in the recurrent layer. For each model, we lesioned a different neuron, and then took the Euclidean distance between each point in $L_i$ and the corresponding point in $Orig$, resulting in a 625 length array of lesion distance values for every lesioned model.

Next, in order to quantify how similar the effects of different lesions were, we calculated the similarity of lesions by correlating the lesion distances from each lesioned model, resulting in an $M \times M$ matrix of similarity values that we refer to as the \emph{lesion similarity} matrix.

    \subsection*{Mouse resting state fMRI data}

All in vivo experiments were conducted in accordance with the Italian law (DL 2006/2014, EU 63/2010, Ministero della Sanitá, Roma) and the recommendations in the Guide for the Care and Use of Laboratory Animals of the National Institutes of Health. Animal research protocols were reviewed and consented by the animal care committee of the Italian Institute of Technology and Italian Ministry of Health. The rsfMRI dataset used in this work consists of $n = 19$ scans in adult male C57BL/6J mice that are publicly available \cite{grandjean2020common,gutierrez2019infraslow}. Animal preparation, image data acquisition, and image data preprocessing for rsfMRI data have been described in greater detail elsewhere \cite{montani2021m1}. Briefly, mice were anesthetized with isoflurane (5\% induction), intubated and artificially ventilated (2\%, surgery). The left femoral artery was cannulated for continuous blood pressure monitoring and terminal arterial blood sampling. At the end of surgery, isoflurane was discontinued and substituted with halothane (0.75\%). Functional data acquisition commenced 45 minutes after isoflurane cessation. Mean arterial blood pressure was recorded throughout imaging sessions. Arterial blood gasses ($paCO_2$ and $paO_2$) were measured at the end of the functional time series to exclude non-physiological conditions. rsfMRI data were acquired on a 7.0-T scanner (Bruker BioSpin, Ettlingen) equipped with BGA-9 gradient set, using a 72-mm birdcage transmit coil, and a four-channel solenoid coil for signal reception. Single-shot BOLD echo planar imaging time series were acquired using an echo planar imaging sequence with the following parameters: repetition time/echo time, 1200/15 ms; flip angle, $30^\circ$; matrix, $100 \times 100$; field of view, $2 \times 2 cm^{2}$; 18 coronal slices; slice thickness, 0.50 mm; 1500 volumes; and a total rsfMRI acquisition time of 30 minutes. Timeseries were despiked, motion corrected, skull stripped and spatially registered to an in-house EPI-based mouse brain template. Denoising and motion correction strategies involved the regression of mean ventricular signal plus 6 motion parameters. The resulting time series were band-pass filtered (0.01-0.1 Hz band) and then spatially smoothed with a Gaussian kernel of 0.5 mm full width at half maximum. After preprocessing, mean regional time-series were extracted for 15314 regions of interest (ROIs) derived from a voxelwise version of  mouse structural connectome \cite{oh2014mesoscale,knox2018high,coletta2020network}.

\subsection*{Mouse Anatomical Connectivity Data}

The mouse anatomical connectivity data used in this work were derived from a voxel-scale model of the mouse connectome made available by the Allen Brain Insitute \cite{oh2014mesoscale,knox2018high} and recently made computationally tractable \cite{coletta2020network}. Briefly, the structural connectome was obtained from imaging enhanced green fluorescent protein (eGFP)–labeled axonal projections derived from 428 viral microinjection experiments, and registered to a common coordinate space \cite{wang2020allen}. Under the assumption that structural connectivity varies smoothly across major brain divisions, the connectivity at each voxel was modeled as a radial basis kernel-weighted average of the projection patterns of nearby injections \cite{knox2018high}. Leveraging the smoothness induced by the interpolation, neighboring voxels were aggregated according to a Voronoi diagram based on Euclidean distance, resulting in a $15314 \times 15314$ whole brain weighted and directed  connectivity matrix \cite{coletta2020network}.

    \subsection*{Human connectomic data}

Structural, diffusion, and functional human brain magnetic resonance imaging (MRI) data was sourced from the Human Connectome Project’s (HCP) young adult cohort (S1200 release). This contained structural MRI (T1w), resting-state functional MRI (rs-fMRI), and diffusion weighted imaging (DWI) data from 1000 adult participants (53.7\% female, mean age = 28.75, standard deviation of age = 3.7, age range = 22-37). A comprehensive report of imaging acquisition and preprocessing is available elsewhere \cite{glasser2013minimal}. In brief, imaging was acquired with a Siemens 3T Skyra scanner with a 32-channel head coil. The rs-fMRI data was collected with a gradient-echo echo-planar imaging (EPI) sequence (run duration = 14:33 min, TR = 720 ms, TE = 33.1 ms, flip angle = 52°, 2-mm isotropic voxel resolution, multi-band factor = 8) with eyes open and instructions to fixate on a cross \cite{smith2013resting}. DWI data was acquired using a spin-echo planar imaging sequence (TR= 5520 ms, TE = 89.5 ms, flip angle = 78°, 1.25 mm iso- tropic voxel resolution, b-values = 1000, 2000, 3000 s/mm2, 90 diffusion weighed volumes for each shell, 18 b = 0 volumes) \cite{uugurbil2013pushing}. HCP minimal processing pipeline was used to preprocess the functional and diffusion imaging data \cite{glasser2013minimal}. In particular, rs-fMRI underwent gradient distortion correction, motion correction, registration to template space, intensity normalization and ICA-FIX noise removal \cite{glasser2013minimal,griffanti2014ica}. The diffusion preprocessing pipeline consisted of b0 intensity normalization, EPI distortion correction, eddy-current-induced distortion correction, registration to native structural space, and skull stripping \cite{glasser2013minimal}.

    \subsubsection*{Diffusion tractography}

    A probabilistic streamline tractography pipeline tailored for the computation of high-resolution human connectomes \cite{tian2021high} was implemented in MRtrix3 \cite{tournier2019mrtrix3} which also adopted recent recommendations detailed elsewhere \cite{tian2021high}. In particular, an unsupervised heuristic was used to estimate macroscopic tissue response functions for white matter (WM), gray matter (GM), and cerebrospinal fluid (CSF) \cite{dhollander2016unsupervised}. Multi-shell, multi-tissue constrained spherical deconvolution was used to estimate fiber orientation distributions (FODs) \cite{jeurissen2014multi}. This information was used to apply combined intensity normalization and bias field correction \cite{dhollander2021multi}. Liberal and conservative brain masks were respectively utilized in the last two steps to mitigate the detrimental effects of an imperfect mask on the respective procedures \cite{tian2021high}. The normalized FODs were used to perform anatomically constrained tractography (ACT) \cite{smith2012anatomically}. A tissue-type segmentation was used to create a mask at the GM-WM boundary to enable brain-wide streamline seeding. Whole-brain tractography was conducted using 2nd-order integration over FODs (iFOD2) \cite{tournier2010improved}. A total of five million streamlines (per participant) were generated that satisfied length (minimum length = 4mm) and ACT constraints.

    \subsubsection*{Human structural connectome}

The high-resolution structural connectivity network was constructed from the whole-brain tractograms \cite{tian2021high}. Notably, streamline endpoints were mapped onto HCP’s CIFTI space comprising two surface meshes of the cortex, in addition to volumetric delineations of several subcortical structures. The surface meshes included a subset of the fs-LR template mesh after removal of the medial wall with respectively 29696 and 29716 vertices for the left and right cortex. In addition, 31870 voxels from the MNI template space were included for subcortical brain regions, resulting in a total of 91282 high-resolution network nodes. Euclidean distance was used to assign each streamline to its closest node pair via nearest neighbor mapping. Streamlines with endpoints falling more than 2mm away from all nodes were discarded. Connectome spatial smoothing with a 6mm FWHM kernel was performed to account for accumulated errors in streamline endpoint location and to increase intersubject reliability of connectomes \cite{seguin2022connectome}. A group-level consensus connectome was constructed by aggregating individual connectivity matrices across all participants. To minimize the pipeline’s computational complexity, group aggregation was applied before smoothing, considering that both aggregation and connectome spatial smoothing entailed linear operations, and the order of implementation does not impact the resulting connectomes.

\hfill

    \subsubsection*{Human thalamocortical connections}
    
We next sought to estimate the thalamocortical structural connectivity connections from the smoothed high-resolution group-level human connectome model. To this end, the cifti structures for the left and right thalamus were combined to form a binary mask of the thalamus. This mask was subsequently used to aggregate group level thalamic connections to the left cortex.

    \subsubsection*{Human functional network} 

The dense group-level functional connectivity network was provided by HCP. Specifically, this network was constructed from a pipeline combining high-resolution individual rs-fMRI data across all participants. First, the minimally preprocessed individual rs-fMRI data were aligned by a multimodal surface matching algorithm (MSMAll) \cite{robinson2014msm}. Next, all timeseries were temporally demeaned followed by a variance normalization \cite{beckmann2004probabilistic} and were passed to a MELODIC’s Incremental Group-PCA \cite{smith2014group}. The group-PCA outputs were renormalized, eigenvalue reweighted, and correlated to create a dense functional connectivity matrix ($91282 \times 91282$). Finally, a subset of this dense connectome was extracted to denote left cortical functional connectivity ($29696 \times 29696$).

\subsection*{Author contributions}

JCT and RFB conceived of experiments, designed analyses, wrote the initial versions of the manuscript, and edited the manuscript. JCT performed all analyses. SM, LC, and AG contributed data. All authors helped edit the manuscript.

\subsection*{Code and data availability}

Code for implementing and training continuous-time recurrent neural networks (CTRNNs) can be found at: \url{https://github.com/gyyang/nn-brain/blob/master/RNN_tutorial.ipynb} Code for the additional analyses will be released upon publication.

For data availability, see the following papers for the mouse data \cite{coletta2020network,gutierrez2019infraslow} and the human structural connectome data \cite{tian2021high}. The dense functional connectivity data from humans is publicly available at: \url{https://db.humanconnectome.org}.

	\bibliography{modular_boundaries_in_RNNs.bib}
	
	\beginsupplement

\section*{Supplementary materials}

In this supplemental section, we describe the results of analyses that support the results and arguments made in our main text.

\subsection*{Representation and selectivity}

\subsubsection{Representation and selectivity in the modules of feed-forward neural networks}\label{sec:feed-forward}

In addition to the analyses in the main results section on the representation and selectivity of modules in recurrent neural networks, we also performed additional analyses on feed-forward neural networks (FFNN). Here, we asked if we could use the information about boundaries between modules in these networks to infer its classification behavior.

\begin{figure*}[!t]
    \centering
    \includegraphics[width=1\textwidth]{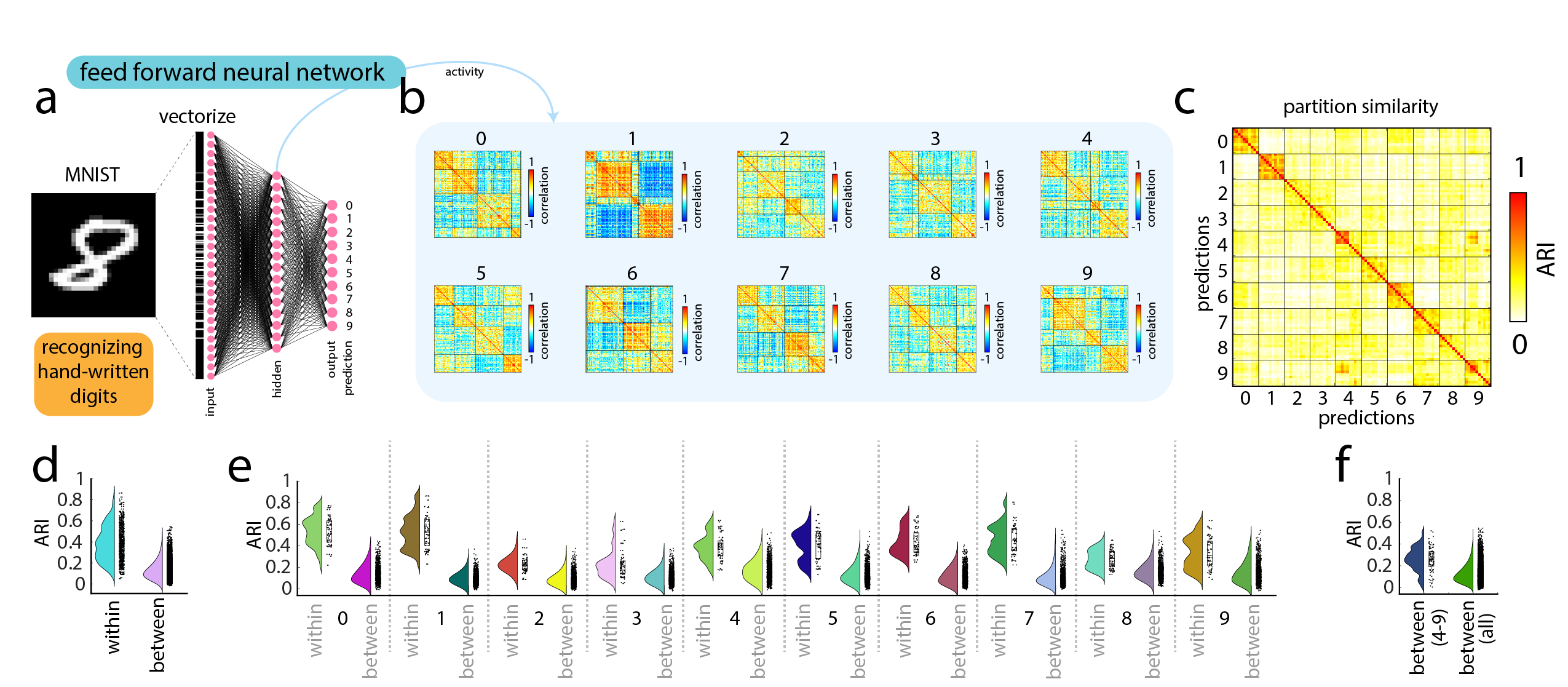}
	\caption{\textbf{modules boundaries are more similar in a feed-forward neural network when the content is similar}, (\emph{a}) Schematic showing how a feed forward neural network recognizes hand-written digits in the MNIST dataset. A vectorized version of the pixels in the image of the handwritten letter is input into an input neurons that project this activity to a hidden layer, and this hidden layers activity is then projected to an output layer where the output neuron with the greatest activity corresponds to the networks decision about the digit it saw in the image. (\emph{b}) Examples of modules found in the correlation matrices of the hidden neurons when this trained network was shown a sample of images from different hand-written digits. (\emph{c}) A similarity matrix comparing the partitions of modules when the network was seeing samples of different digits. Note how the similarity of the partitions is higher when the neural network is seeing the same digit (different sample) then when seeing a different digit. (\emph{d}) Boxplots comparing within digit and between digit similarity values. (\emph{e}) Boxplots comparing within digit and between digit similarity values separately for each digit. (\emph{f}) Boxplots comparing the similarity values between digit 4 and digit 9, with the similarity values between all other digits.} 
	\label{mnist}
\end{figure*}

First, we trained a simple FFNN to recognize hand-written digits (MNIST dataset; Fig. \ref{mnist}a). We then investigated the activity in the hidden layer of a FFNN, and we found modules. Following training, we divided the training data into the different digits (0 through 9; 1,000 samples each), and further divided the data for each of these digits into ten separate datasets (100 samples each). We produced activity in these networks by presenting them with the examples in these datasets. We then found moduleboundaries for each of these datasets. We then compared the boundaries from each of these datasets (10 sets of 100 samples per digit) with one another using adjusted Rand index (ARI). We found that partition labels (representing the boundaries) were more similar when the network was recognizing the same digit than when it was recognizing different digits (Fig. \ref{mnist}a-c; two-sample \emph{t}-test, all $p < 10^{-15}$). Interestingly, we also see that the greatest similarity between the presentation of two digits is between hand-written samples of the number 4 and the number 9. We note that in their handwritten form, these numbers look very similar.

Next, we performed a similar analysis on a large language model. More specifically, we used a deep transformer model from OpenAI (GPT-2). Similar to the previous analysis, we wished to see if semantically similar information resulted in similar boundaries between modules. 

We provided the network with nine semantically related sentences. The first three sentences were related to cars. The second three were related to houses, and the third three were related to cats. Additionally, these three sets were related in the following way. Each set contained sentences with the following form: \emph{a)} What is a \emph{x}? \emph{b)} Where is the \emph{x}?
\emph{c)} This is a \emph{x}.
\emph{x} is either cars, houses, or cats. So, these nine sentences had relationships of both subject-matter and form. 

\begin{figure*}[!t]
    \centering
    \includegraphics[width=1\textwidth]{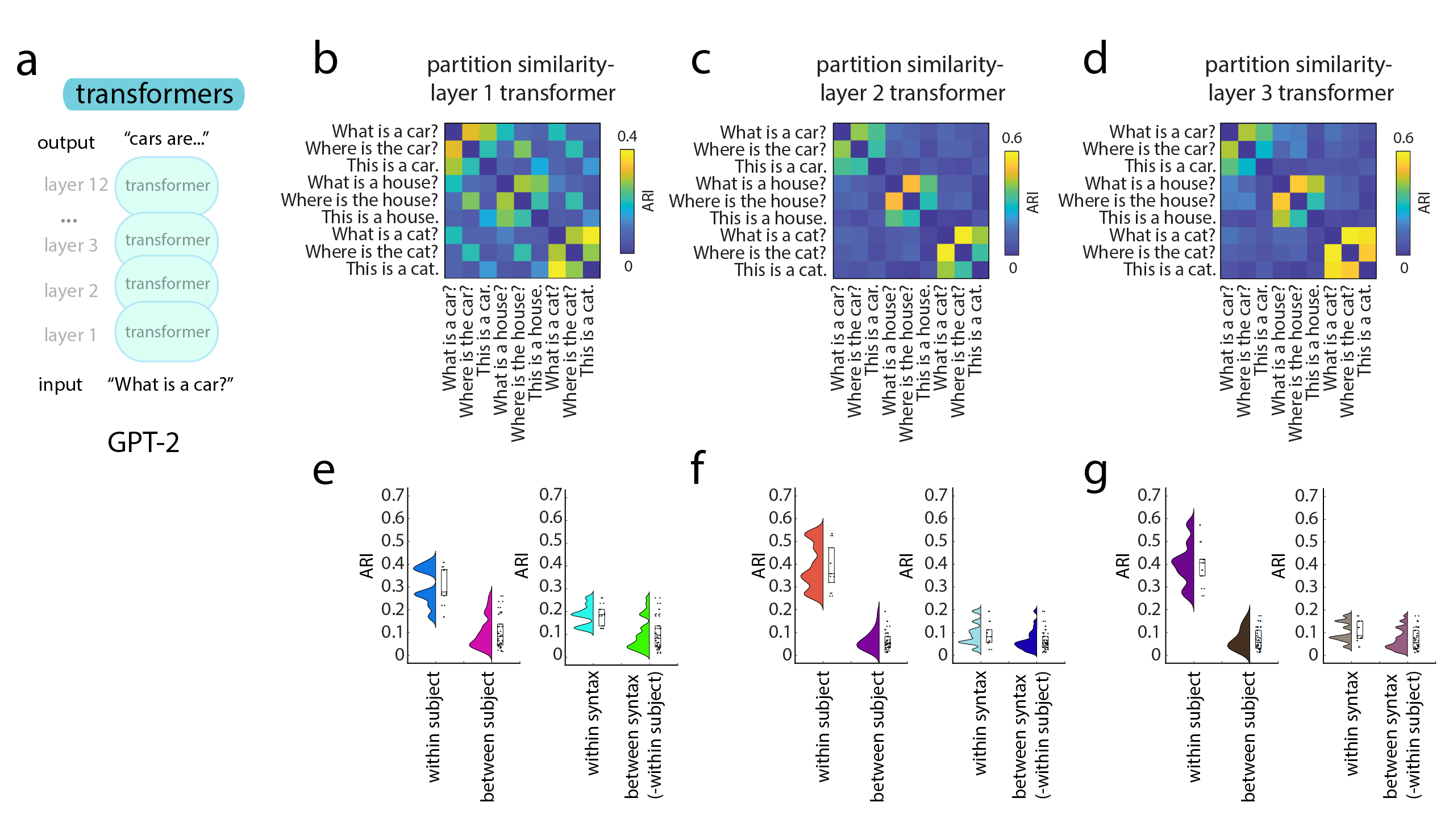}
	\caption{\textbf{modules boundaries are more similar in a large language model when the content is similar}, (\emph{a}) Schematic representation of a deep transformer model with 12 layers, outfited with pretrained weights from GPT-2. (\emph{b}) Partition similarity of modules in the first layer, second layer (\emph{c}) and third layer (\emph{d}) of the transformer model when asked different semantically related questions. (\emph{e/f/g}) First panel shows boxplots comparing within subject (``car",``house",``cat") versus between subject similarity values for layer 1/2/3. Second panel shows boxplots comparing within syntax (``what is a x?", ``where is the x?", ``This is a x") versus between syntax similarity values for layer 1/2/3. Within subject similarity values were removed from consideration in the between syntax boxplot. } 
	\label{transformer}
\end{figure*}

We found that the first layer of the transformer model had modules that were related to both subject-matter and form, but the moduleboundaries in the following two layers only represented subject-matter. More specifically, in the first layer both the partitions with the same subject-matter (cars, houses, cats) were highly similar, and the partitions with the same form were highly related (Fig. \ref{transformer}b,e; within subject $p < 10^{-15}$, within syntax $p < 10^{-6}$). But in layer's 2 and 3, the partitions were mainly similar when the subject-matter was similar (Fig. \ref{transformer}c,d,f,g; layer 2: within subject $p < 10^{-15}$, within syntax $p = 0.02$, layer 3: within subject $p < 10^{-15}$, within syntax $p = 0.001$).

Taken together, these results suggest that the boundaries between modules in artificial neural networks hold information for task-relevant functions.

\subsection*{Origin}

\subsubsection{Generative models of module boundaries in feed-forward neural networks}\label{sec:generative}

We found that we could assign the output neurons to modules based on two properties of the input weights: 1) the sign of the weights from either input neuron or 2) which of the two input neurons it gave the greatest weight. 

In order to tease apart the contributions of these two types of asymmetries to the development of moduleboundaries, we developed generative models that isolate one or the other. Generative models are particularly useful in this case as they allow us to see how modules boundaries change as we change a parameter of the model. Using generative models, we show that these two basic conditions create modules in output neurons. 

\begin{figure*}[!t]
    \centering
    \includegraphics[width=1\textwidth]{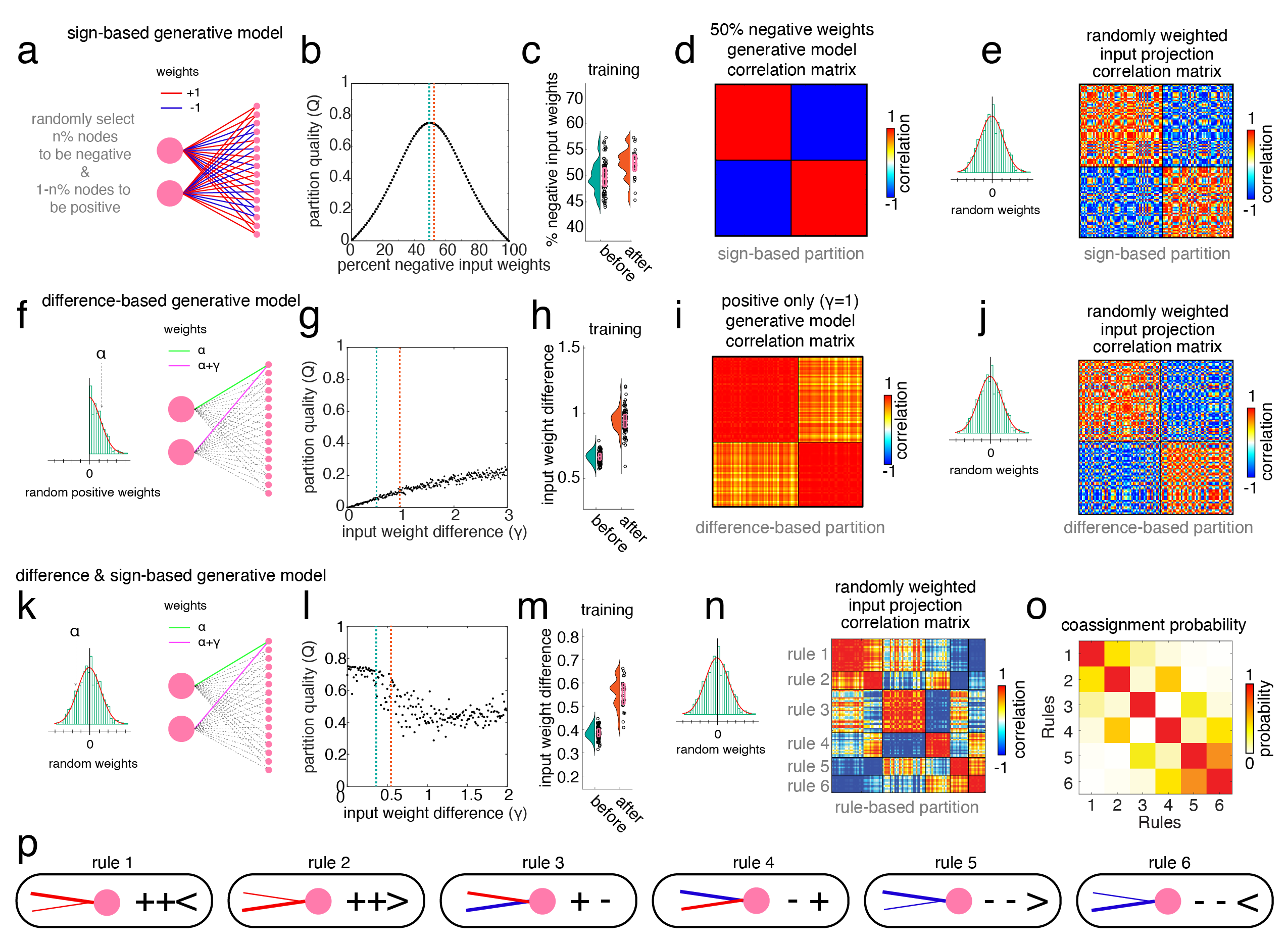}
	\caption{\textbf{Generative models of modules boundaries}, (\emph{a}) Schematic of the sign-based generative model. The weights to each output neuron are given one of two values [+1,-1]. We randomly select n\% of neurons to have weights of -1 and 1-n\% of neurons to have weights of +1. (\emph{b}) As we increased the n\% of neurons with -1 weights partition quality ($Q$) increases from a value of zero until it peaks when 50\% of neurons have weights of -1. Dashed lines are color coded to correspond to boxplots in the next panel. (\emph{c}) Boxplots showing the \% of negative weights in the input connection weights of 100 RNN models before and after training. (\emph{d}) Correlation matrix for output neurons with the sign-based generative model when reorganized into two clusters according to the sign of weights converging on an output neuron. (\emph{e}) Correlation matrix from a random input connection matrix reordered by the sign-based partition. (\emph{f}) Schematic of the difference-based generative model. For each pair of input connections to an output neuron a connection is chosen at random and given a positive weight $\alpha$ from the absolute value of a Gaussian distibution centered on zero. The other connection is given the value $\alpha + \gamma$. (\emph{g}) As we increased $\gamma$ the partition quality ($Q$) increased from a value of zero. Dashed lines are color coded to correspond to the next panel. (\emph{h}) Boxplots showing the weight difference between two input neurons in the input projections of 100 RNN models before and after training. (\emph{i}) Correlation matrix of the difference-based generative model when reorganized into two clusters according to the input neuron with the most weight (difference-based partition). (\emph{j}) Correlation matrix from a random input connection matrix reordered by the difference-based partition. (\emph{k}) Schematic of the difference \& sign-based generative model. This is the same as the difference-based model, but negative weights are allowed. (\emph{l}) As we increased $\gamma$ the partition quality ($Q$) changed. Dashed lines are color coded to correspond to the next panel. (\emph{m}) Boxplots showing the weight difference between two input neurons in the input connections of 100 RNN models before and after training. (\emph{n}) Correlation matrix from a random input connection matrix reordered by the rule-based partition. (\emph{o}) Mean probability of the coassignment of different rule based neurons to the same modules when running modularity maximization with the Louvain algorithm 1000 times. (\emph{p}) Schematic of the rules for the rule-based partition.} 
	\label{input_projections}
\end{figure*}

First, modules were created when the weights from input neurons were composed of both positive and negative weights. For our generative model, we randomly assigned weights to either 1, or -1. We found that as you modulate the percentage of weights that are negative, you increase the quality ($Q$) of the partitions of output activity into modules (Fig. \ref{input_projections}a,b). $Q$ is at a value of zero when all weights have a value of 1, and gradually increases as negative weights (-1) are added to the connections. $Q$ peaks when 50\% of the weights are negative and the modules are equally sized.

Intuitively, when the sign of connections converging on an output neuron is mostly negative, then input activity causes the output neuron to decrease its activity. Conversely, if the sign is mostly positive, input activity will cause the output neuron to increase its activity. This means that when both positive and negative weights are found in the input weights, neurons with the same sign will produce activity that is positively correlated, while neurons with different signs will produce activity that is negatively correlated. This divides the correlation matrix into modules based on the sign of these weights (Fig. \ref{input_projections}d,e).

The second condition that produces modules is when there is a difference in the weight of the $N$ connections converging on each output neuron. For our generative model, we randomly selected each weight from a Gaussian distribution centered on zero. In order to differentiate this model from the previous model that was based on the sign of the weights, we took the absolute value of these weights (Fig. \ref{input_projections}f). We then assigned one of the two input connection weights to this value. The second connection weight received the same weight plus some value $\gamma$ that defined the difference between the weights. We found that if all of the weights were positive and there was no difference between the weights ($\gamma = 0$), then the input created no modules in the output activity, however as we increased $\gamma$ the quality ($Q$) of modules partitions also increased (Fig. \ref{input_projections}g).

Intuitively, when the sign of these weights is the same, but the weight values converging on an output neuron differ, then the output neuron will produce more activity when one input neuron is active than when the other input neuron is active. This results in some output neurons activating more in the presence of one input neuron, and some output neurons activating more in the presence of the other input neuron. Therefore the activity of output neurons in the first group will be more positively correlated to one another than to the activity of output neurons in the next group, thereby producing modules (Fig. \ref{input_projections}i,j).

We also initialized and trained 100 RNN models on a perceptual decision-making task in order to identify what values (\% of negative weights and input weight differences $\gamma$) these input connection weights take on before and after training, and we found that the \% of negative weights as well as the input weight differences increase across training (Fig. \ref{generative}a,d; two-sample \emph{t}-test, $p < 10^{-15}$). 

We created a third generative model that allowed for signed weights as well as input weight differences. To do this, we randomly initialized the weight from one of two input connections converging to a value ($\alpha$). The weight of the second connection was assigned to $\alpha - \gamma$ (Fig. \ref{input_projections}k). We found that while the quality $Q$ of the partition of activity into modules is modulated by $\gamma$ it still maintains a high value (Fig. \ref{input_projections}n). 

modules formed by the interaction between these two conditions in the input connections can be neatly partitioned using six (or eight) rules that define different combinations of these conditions. Each rule depends upon the sign and/or weight difference between the two connections converging on each output neuron (Fig. \ref{input_projections}p). The rules are as follows: \emph{1)} connection $A$ and $B$ are positive, but connection $A$ is stronger, \emph{2)} connection $A$ and $B$ are positive, but connection $B$ is stronger, \emph{3)} connection $A$ is positive and connection $B$ is negative, \emph{4)} connection $A$ is negative and connection $B$ is positive, \emph{5)} connection $A$ and $B$ are negative, but connection $A$ is stronger, \emph{6)} connection $A$ and $B$ are negative, but connection $B$ is stronger. These 6 rules can become 8 rules if you sub-divide rules 3 and 4 based on the connection with the greatest absolute weight value.

\begin{figure*}[!t]
    \centering
    \includegraphics[width=1\textwidth]{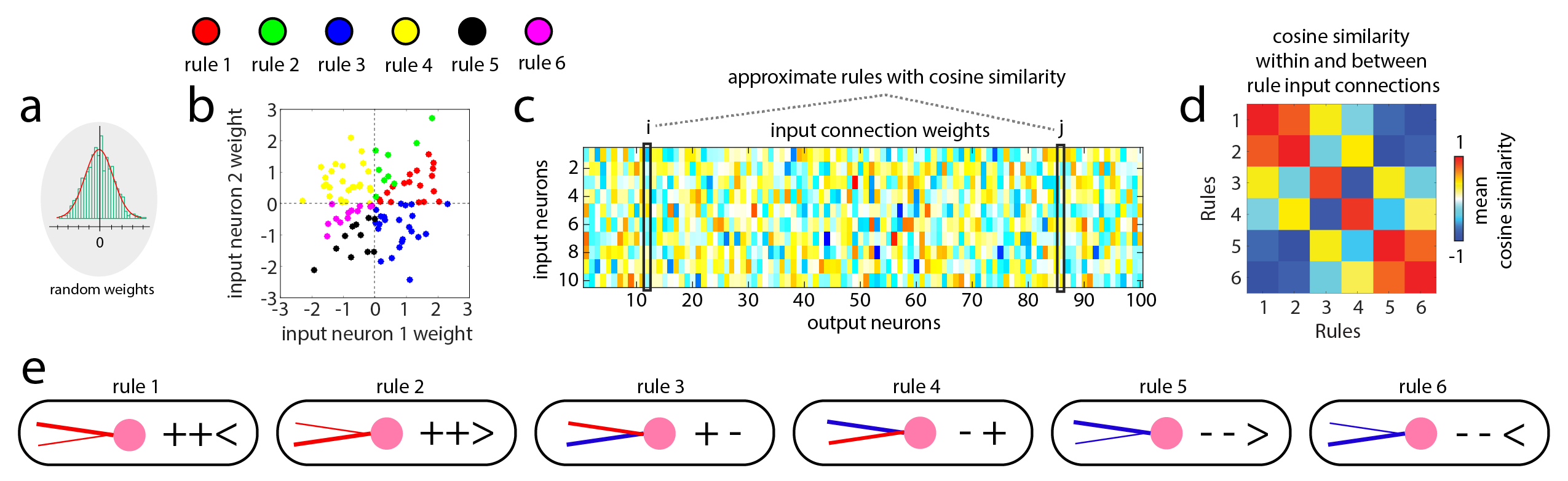}
	\caption{\textbf{Cosine similarity approximates rule-based partition}, (\emph{a}) Schematic showing that the weights considered in this figure were randomly drawn from a Gaussian distribution centered on zero. (\emph{b}) Plot of each output neurons input weights color coded by the rule that the output neuron belongs to. Note how the rules cluster. The angle between any two points within a cluster tends to be small. (\emph{c}) Example of an input connection weight matrix with 10 input neurons and 100 output neurons. Cosine similarity is calculated between the input weights of all i by j output neurons. (\emph{d}) Matrix showing the mean cosine similarity between different output neurons. This matrix is organized by the rule that the two output neurons belonged two. The strong values along the diagonal indicate that output neurons with the same rule are more likely to have high cosine similarity. } 
	\label{cosine}
\end{figure*}

Importantly, these rules are only well-defined with two input neurons. If we would like to partition activity based on the structure of input connections with $N$ neurons, we need some way of approximating this for $N$-dimensional input. As such, we approximated these rules by cosine similarity. More specifically, we can approximate the partitions produced in the correlation matrix of output activity by clustering a matrix of the pairwise cosine similarity between the in-weights of all output neurons (Fig. \ref{cosine}d).

\subsubsection{Input neuron weights guide the development recurrent neuron function}\label{sec:initial}

In the main results section, we showed that input neuron connection weights influence the structure of modules that form in the recurrent neurons of RNNs and we see a similar result in brains (mouse and human). We also showed that, after training, modules are selective for different task informaton. Inspired by these results, in this section we ask if input connection weights bias the development of specialized modules.

Here, we show that if you initialize these RNNs with the same input connection weights on multiple training runs, then the modules across all of these training runs are highly similar. Additionally, we show that if you initialize the input connection weights and the output connection weights to be identical, then the modules after training are even more similar across runs, and the strength of the weights in the recurrent layer is also highly similar. This suggests that the modules of recurrent neurons will slightly alter themselves to conform to the output neuron weights, and so by fixing the output neuron weights, this conformation no longer needs to occur.

\begin{figure*}[!t]
    \centering
    \includegraphics[width=1\textwidth]{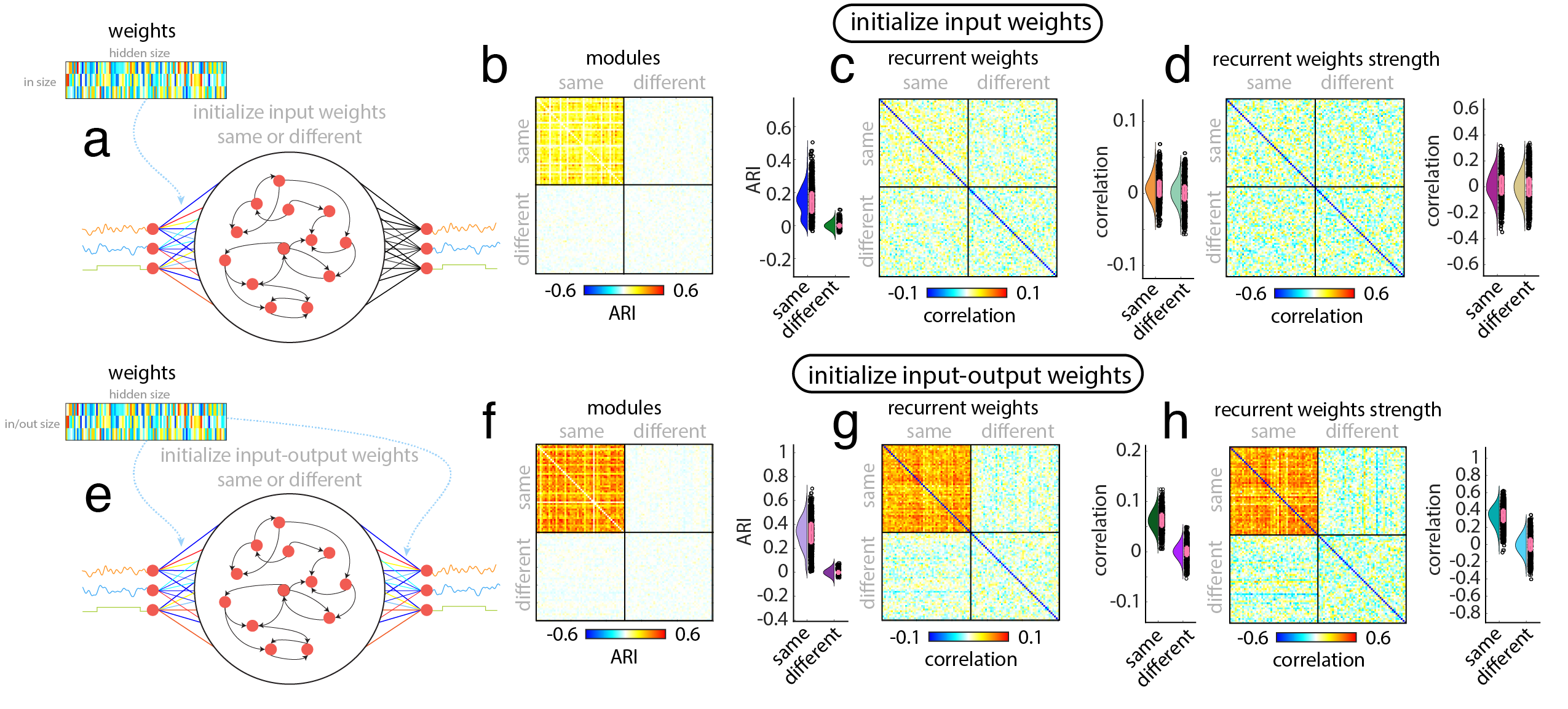}
	\caption{\textbf{Initial input weights guide network structure and function across training}, (\emph{a}) Schematic showing the condition where we initialize the input weights for the RNN to be the same when training 50 different RNNs. (\emph{b}) Similarity matrix showing the similarity (adjusted Rand index) of the modular partitions for 50 RNNs where the input weights are initialized the same and 50 RNNs where they are initialized randomly. Partitons were found using modularity maximization with the Louvain algorithm on the correlation matrix of recurrent activity. Also, boxplots showing the same information. (\emph{c}) Same as b, but comparing similarity of weights connecting recurrent neurons. (\emph{d}) Same as b, but comparing similarity of the strength of weights connecting recurrent neurons. (\emph{e}) Schematic showing the condition where we initialize the input weights and the output weights to be the same set of weights $\gamma$. $\gamma$ is the same across 50 different RNNs. (\emph{f}) Same as b, but for the modules in the condition where input and output weights were matched. (\emph{g}) Same as b, but comparing similarity of weights connecting recurrent neurons. (\emph{h}) Same as b, but comparing similarity of the strength of the weights connecting recurrent neurons. } 
	\label{same_diff}
\end{figure*}

To run this analysis, we trained 50 RNNs and initialized the weights of the input connections to be the \emph{same} before training (Fig \ref{same_diff}a). We also trained 50 RNNs where these initial weights were randomly initialized (\emph{different}). After training, we compared these networks in terms of three features: the partition of the correlation matrix into modules, the recurrent weights, and the strength of the recurrent weights. We found that RNNs where the weights of input connections were initialized with the same values had significantly more similar partitions (Fig. \ref{same_diff}b; two-sample \emph{t}-test; $p < 10 ^{-15}$). We also tested how similar the recurrent weights and recurrent weights strength were across these 100 RNNs and found a very small effect for the recurrent weights, indicating that they were very slightly more similar when the input connection weights had been initialized with the same values (Fig. \ref{same_diff}c; two-sample \emph{t}-test; $p < 10 ^{-15}$). 

Next, we performed a similar analysis with RNNs where the weights of the input connections and the output connections were initialized to be the same values. We trained 50 RNNs and initialized the input connection weights with a set of weights $\gamma$ and initialized the output layer with the same weights $\gamma$. We kept $\gamma$ the same on the initialization of all 50 RNNs. We also trained an additional 50 RNNs where the weights of the input layer and the output layer were randomly initialized. We found that the modules boundaries of the RNNs were more similar when the input and output connection weights were initialized to be the same (Fig. \ref{same_diff}f; two-sample \emph{t}-test; $p < 10 ^{-15}$). Furthermore, we also found that the weights of the recurrent neurons are significantly more similar in the input-output matched condition whether comparing the weights individual or the strength of the weights in the recurrent neurons(Fig. \ref{same_diff}g-h; two-sample \emph{t}-test; $p < 10 ^{-15}$).

These results suggest that the boundaries between modules in RNNs is related to the weights of the input connections that are provided to the network before training. This might have interesting implications, given that these weights are normally random.

	\clearpage

 \end{document}